\documentclass[sigconf]{acmart}

\AtBeginDocument{%
  }

\copyrightyear{2026}
\acmYear{2026}
\setcopyright{cc}
\setcctype{by}
\acmConference[CHI '26]{Proceedings of the 2026 CHI Conference on Human Factors in Computing Systems}{April 13--17, 2026}{Barcelona, Spain}
\acmBooktitle{Proceedings of the 2026 CHI Conference on Human Factors in Computing Systems (CHI '26), April 13--17, 2026, Barcelona, Spain}
\acmPrice{}
\acmDOI{10.1145/3772318.3790472}
\acmISBN{979-8-4007-2278-3/2026/04}

\graphicspath{{Graphics/}}
\usepackage{makecell}

\usepackage{graphicx}
\usepackage{color}
\usepackage{enumitem}
\usepackage{booktabs}
\usepackage{multirow}
\usepackage{xcolor,soul}
\usepackage{subcaption} 
\usepackage{longtable}
\usepackage{array}
\usepackage{epstopdf}

\graphicspath{{Graphics/}}

\newcommand{\Npre}{97}
\newcommand{\NQApre}{47}
\newcommand{\N}{44}

\newcolumntype{L}[1]{>{\raggedright\arraybackslash}p{#1}}

\begin{document}

\title{Supporting Informed Self-Disclosure: Design Recommendations for Presenting AI-Estimates of Privacy Risks to Users}

\author{Isadora Krsek}
\email{ikrsek@andrew.cmu.edu}
\orcid{0000-0001-9624-5077}
\affiliation{%
  \institution{Carnegie Mellon University}
  \city{Pittsburgh}
  \state{PA}
  \country{USA}
}

\author{Meryl Ye}
\email{ye2@andrew.cmu.edu}
\orcid{0000-0001-8215-9020}
\affiliation{%
  \institution{Carnegie Mellon University}
  \city{Pittsburgh}
  \state{PA}
  \country{USA}
}

\author{Wei Xu}
\email{wei.xu@cc.gatech.edu}
\affiliation{%
  \institution{Georgia Institute of Technology}
  \city{Atlanta}
  \state{Georgia}
  \country{USA}
}

\author{Alan Ritter}
\email{ritter.alan@gmail.com}
\affiliation{%
  \institution{Georgia Institute of Technology}
  \city{Atlanta}
  \state{Georgia}
  \country{USA}
}

\author{Laura A. Dabbish}
\email{dabbish@cmu.edu}
\affiliation{%
  \institution{Carnegie Mellon University}
  \city{Pittsburgh}
  \state{Pennsylvania}
  \country{USA}
}

\author{Sauvik Das}
\email{sauvik@cmu.edu}
\affiliation{%
  \institution{Carnegie Mellon University}
  \city{Pittsburgh}
  \state{Pennsylvania}
  \country{USA}
}

\renewcommand{\shortauthors}{Krsek et al.}

\begin{abstract}
People candidly discuss sensitive topics online under the perceived safety of anonymity; yet, for many, this perceived safety is tenuous, as miscalibrated risk perceptions can lead to over-disclosure. Recent advances in Natural Language Processing (NLP) afford an unprecedented opportunity to present users with quantified disclosure-based re-identification risk --- i.e., ``population risk estimates'' (PREs). How can PREs be presented to users in a way that promotes informed decision-making, mitigating risk without encouraging unnecessary self-censorship? Using design fictions and comic-boarding, we story-boarded five design concepts for presenting PREs to users and evaluated them through an online survey with $N={\N}$ Reddit users. We found participants had detailed conceptions of how PREs may impact risk awareness and motivation, but envisioned needing additional context and support to effectively interpret and act on risks. We distill our findings into four key design recommendations for how best to present users with quantified privacy risks to support informed disclosure decision-making.
\end{abstract}



\begin{teaserfigure}
\centering
    \includegraphics[width=14cm]{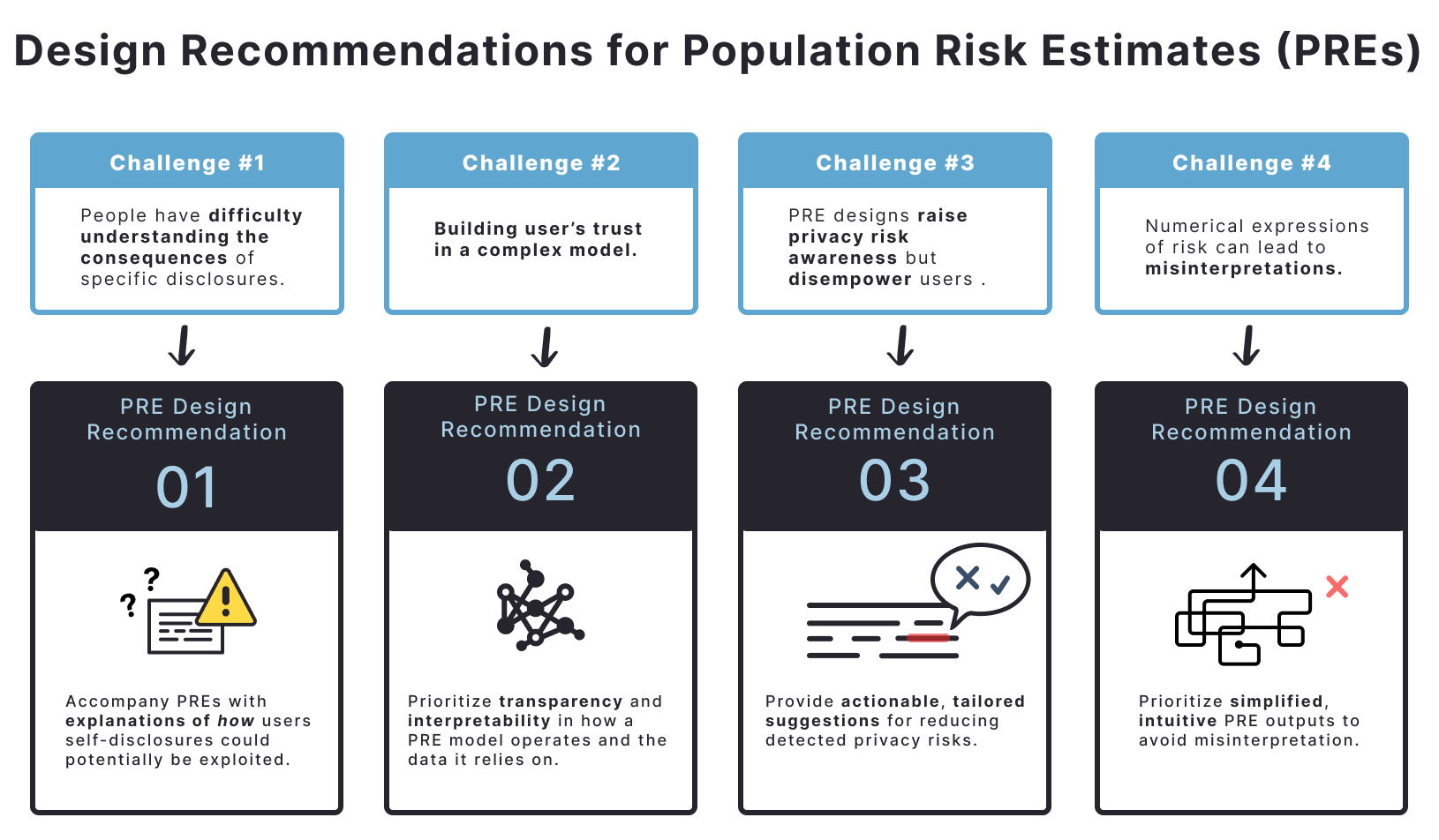}
    \caption{Presentation of the challenges that arose in our findings, and the four design recommendations that emerged from our work around how population risk estimates (PREs) should be presented in order to maintain user engagement, while promoting informed decision making on potential privacy risks stemming from self-disclosure.}
        \Description{Image depicting the challenges that emerged from our work, and how they directly tie to each design recommendation. Challenges are listed out in a row at the top, with the corresponding design challenges directly below. Challenge \#1 was that people have difficulty understanding the consequences of specific disclosures, and the corresponding design recommendation is that PREs should be accompanied with an explanation of how disclosures may be exploited. Challenge \#2 is building trust in complex models, and the corresponding design recommendation is to prioritize transparency and interpretability in how a PRE model operates and the data it relies on. Challenge \#3 is that PRE designs raise privacy risk awareness but may disempower users, and the corresponding design recommendation is that PREs should be accompanied with actionable, tailored advice for reducing detected privacy risks. Finally, challenge \#4 is that numerical expressions of risk can lead to misinterpretations, as such the final design recommendation is to prioritize simplified intuitive PRE outputs to avoid misinterpretation.}
    \label{fig:teaser}
\end{teaserfigure}

\maketitle

\section{Introduction}
The promise of pseudonymity in online fora prompts millions of users to seek informational and emotional support for deeply personal matters. For example, people use Reddit to seek advice on topics ranging from mental health challenges to relationship difficulties, serving as crucial resources for individuals dealing with sensitive topics that might be difficult to discuss in face-to-face interactions \cite{Choudhury2014MentalHD, de2014mental, henninger2020gave}. 

While users want to access informational and emotional support, they do so pseudonymously because there may be a perceived risk of harm if their identities are attached to the support they are seeking \cite{kang2013people, zhang2018health}. However, the safety boundaries of pseudonymity are ambiguous and abstract: to date, users have had few mechanisms to reason about how likely they are to be re-identified from disclosing personal details in isolation or in tandem, such as their age, gender, location, and other identifying characteristics. This digital vulnerability is often accompanied by significant privacy risks that are abstract and difficult for users to comprehend. While users post online self-disclosures with the intention of reaping social benefits, they may forget or lack understanding of the potential re-identification harms---including stalking, identity theft, and blackmail \cite{acquisti2006imagined, wang2011regretted, kramer2020mastering}. The asymmetry between tangible benefits and abstract risks creates a defining problem: How can we help users reap the social benefits of online self-disclosure while mitigating the privacy risks that those disclosures entail? 

Recent advances in language model reasoning suggest that it is now possible to provide users with quantifiable population risk estimates (PREs)---statistical measures that indicate how many people in a population share identifying characteristics and thus how uniquely identifiable a user might be \cite{zheng2025probabilistic}. For example, disclosing that one is a woman in the U.S. might identify a user as being one of about $k=170,000,000$ people (half of the U.S. population). Disclosing that one is a woman in the U.S. who plays tennis might reduce that $k$-value to around $k=11,000,000$ \cite{USTA_2024}. Each subsequent disclosure may reduce this $k$-value, in turn (e.g., that one lives in Arkansas, is 22 years old, and plays tennis at a specific court). These estimates might be able to help make the abstract safety risks of self-disclosure more concrete. However, prior work also suggests that while users find themselves aware of potential risks, they struggle with knowing how to act on this information \cite{krsek2025measuring}. Therefore, we hypothesize that it is not enough to simply present PREs to users, but to identify the appropriate way to present these values to users to promote informed decision-making. 

We present a speculative design exploration of how PREs can be effectively communicated to users making real-time sharing decisions. Following the Security and Privacy Acceptance Framework \cite{das2022security}, which identifies awareness, motivation, and ability as critical barriers to end-user acceptance of privacy technologies, we investigate four key research questions:

\begin{itemize}
\item {\texttt{RQ1}}: How do PREs influence users’ \textbf{awareness} of the risks associated with individual disclosures?
\item {\texttt{RQ2}}: How do PREs impact users’ perceived \textbf{motivation} to address self-disclosure risks?
\item {\texttt{RQ3}}: How do PREs affect perceived \textbf{ability} to address self-disclosure risks?
\item {\texttt{RQ4}}: How might we present PREs to best help users make informed disclosure decisions, balancing both their sharing needs and privacy concerns?
\end{itemize}

Using comic-boards that situate different risk estimation designs within illustrated scenarios, we report on results from an exploratory design fiction study conducted with {\N} Reddit users recruited from Prolific. Recognizing the challenge of soliciting meaningful feedback on unfamiliar privacy technologies, we employed a speed-dating methodology where different approaches to presenting PREs were embedded within illustrated comic-boards (Fig. \ref{fig:fullcomicboard}). Participants evaluated these scenarios and reflected on how the depicted risk estimation tools might affect user behavior, decision-making, and affective response. This approach allowed us to capture nuanced reactions to various interface designs while grounding abstract privacy concepts in relatable user scenarios. Additionally, by operationalizing our study within Reddit, an online platform with rich text-based disclosure norms across many communities, we were able to examine how PREs might apply across a variety of support-seeking scenarios inspired by prior literature \cite{kang2013people}.

Our findings show that PREs were generally envisioned to raise people's awareness of disclosure risks and improve their ability to address those risks prior to posting. PREs effectively heightened risk perception in 74\% of the envisioned outcomes participants articulated, with most participants envisioning characters accepting and meaningfully engaging with quantified privacy feedback in their reflections. PREs also motivated primarily adaptive responses---with a majority of participants describing moderate editing behaviors that preserved communicative intent. Participants also envisioned characters successfully modifying posts to reduce risk, with 79 of 132 reflections ending in characters' successful evasion of re-identification threats.

Yet, these benefits were accompanied by implementation challenges that should be addressed prior to deployment. In analyzing the envisioned complications and challenges characters faced, we distilled four key design recommendations for how to present PREs to maximize their utility in privacy decision-making (see Fig. ~\ref{fig:teaser}): PREs must (i) be accompanied with actionable suggestions for preserving communicative intent while reducing risk or alternative methods to seek support when the risk is too high; (ii) explain how the value of the PRE was determined, with plausible ways attackers might exploit these disclosures; (iii) communicate risks in a way that promotes careful behavior without causing users to censor themselves unnecessarily; and, (iv) use clear, interpretable language and visuals that avoid technical jargon and misinterpretation. While our comic-boards focused on privacy harms from online self-disclosure, these recommendations, grounded in our systematic analysis of participant feedback, provide a rich foundation for the design of AI-estimates of privacy risks for users at large. 

This work makes the following key contributions to the design and deployment of population risk estimate tools:
\begin{itemize}
    \item A systematic examination of how different presentations of PREs affect user awareness, motivation, and perceived ability to manage disclosure risks in online contexts.
    \item Behavioral insights into the psychological and emotional impacts of quantified privacy feedback, including identification of design patterns that can lead to counterproductive outcomes such as increased anxiety and self-censorship.
    \item Evidence-based design recommendations for presenting PREs that maintain user engagement while promoting informed privacy decision-making.
\end{itemize}

\section{Related Works}

\subsection{Online Self-Disclosure \& Anonymity}
Online self-disclosure has become an integral aspect of digital social interaction. Research has demonstrated that self-disclosure establishes solidarity and community, builds new relationships between users, and provides users with increased confidence and a sense of belonging \cite{andalibi2019announcing, lampe2007familiar}. In online social networks, users are primarily motivated to disclose information for the convenience of maintaining and developing relationships and other platform enjoyment, though privacy risks represent a critical barrier to information disclosure \cite{krasnova2010online}.

The appeal of pseudonymous platforms lies in their perceived ability to provide the benefits of self-disclosure while mitigating social risks. \citet{kang2013people} interviewed 44 people across multiple continents to understand their motivations for seeking anonymity, finding a wide variation in people's experiences and life situations leading them to seek online anonymity. They identified several categories of personal threat models threats that users were concerned about: known others (family members, employers, teachers, and former partners), organizations (government and business entities that could reuse or punish disclosed information), other users within the same community or platform, and unknown malicious entities lurking online such as online predators (criminals, hackers, scammers, stalkers), \cite{kang2013people}. These threat models informed the design of the comic-boards used in our study. 

Furthermore, this research revealed diverse motivations ranging from protection against social stigma to professional consequences, with users employing various technical and behavioral strategies to achieve desired levels of anonymity. A 2013 Pew Research survey found that 59\% of Americans believed people should have the ability to use the internet completely anonymously, with 86\% of internet users having taken steps to mask their digital footprints \cite{rainie2013anonymity}. These studies show that users' motivations for anonymous self-disclosure are contextual and personal.

Despite users' efforts to maintain anonymity, research has consistently demonstrated the fragility of anonymization techniques and the ease of re-identification attacks. Prior work has shown that remarkably little information is needed for re-identification: just three demographic attributes can identify approximately 83\% of Americans, while 15 attributes can identify 99.98\% \cite{montjoye2013unique}. Even seemingly innocuous combinations like age, gender, and a medical diagnosis can be sufficient to re-identify individuals \cite{sweeney2000simple}. The combination of date of birth, gender, and zip code is enough to uniquely identify at least 87\% of the US population in publicly accessible databases.

These findings reveal a disconnect between users' intuitive understanding of privacy risks and the realities of re-identification. Information that appears harmless in isolation can become highly identifying when combined with other data points or external datasets, creating a ``privacy paradox''---where users' stated privacy concerns don't align with their disclosure behaviors. Users' understanding of privacy risk is further guided by their sense of anonymity. Higher perceptions of anonymity can result in ``benign disinhibition'' --- i.e., disclosing more personal information because one feels more secure and less identifiable \cite{lapidot2015benign}. As such, re-identification risks are especially relevant to users of pseudonymous online communities, like Reddit, where perceived anonymity leaves users prone to benign disinhibition effects. For this reason, we focus our study on Reddit, recruiting users of the pseudonymous online platform so as to elicit feedback from those users who would be most likely to benefit from PREs in their day-to-day activities.

\subsection{Designing for Risk Awareness}
The challenge of raising privacy risk awareness has been a longstanding interest in security and privacy literature. \citet{schaub2015design}'s design space for effective privacy notices identified multiple dimensions for privacy communication, including timing, channel, control, and modality, emphasizing that effective privacy notices must be contextual, actionable, and comprehensible to users. Their framework established that privacy information must be presented at the right moment, through appropriate channels, with meaningful user control, to support informed decision-making. Building on this foundation, \citet{acquistiNudges2017}'s comprehensive review of nudging for privacy and security identified problems users face in privacy decision-making, including incomplete information, bounded rationality, and cognitive biases that lead to decisions users may later regret. Their work provides a taxonomy of nudging approaches (e.g., presentation nudges, information nudges, defaults, timing, and social influence) while establishing a framework for designing privacy nudges that maintain user autonomy.

However, it's important to note that risk awareness alone does not guarantee action. Traditional approaches to privacy protection, such as granular privacy settings, have shown limited effectiveness and may even paradoxically increase disclosure behavior, as users interpret greater control as reduced risk \cite{brandimarte2013misplaced}. Empirical studies have demonstrated both the potential and limitations of risk communication tools. \citet{ur2012does}'s foundational work on password meters showed that real-time feedback can influence user behavior, but the design of these feedback mechanisms critically affects their effectiveness. Their findings revealed that stringency of scoring matters more than visual appearance--meters requiring high estimated guesses led to significantly stronger passwords, while lenient meters showed minimal security improvements. Later work demonstrated that detailed text feedback led to 44\% improvement in password strength over basic policies, establishing the design principle of providing specific, actionable feedback rather than just colored indicators \cite{ur2017design}. Similarly, Wang et al.'s research on Facebook privacy interfaces revealed the complexities of supporting user privacy decision-making in social media contexts \cite{wang2011regretted, wang2014field}. Their work on privacy regrets identified seven primary causes of regrettable posts, including unintended audience exposure and misjudging social norms, providing foundational understanding of privacy risks that informed subsequent interface design work. Field trials of privacy nudges demonstrated that audience reminders effectively prevented unintended disclosures through lightweight, non-intrusive awareness tools, showing that interface modifications can influence privacy decision-making without major user burden.

These empirical findings have informed broader frameworks for designing risk awareness tools. The Security, Privacy, Awareness, and Feedback (SPAF) framework provides a structured approach by identifying three critical barriers preventing adoption of expert-recommended security behaviors: Awareness (understanding threats and mitigation strategies), Motivation (willingness to employ practices), and Ability (correctly implementing measures) \cite{das2022security}. The SPAF framework emphasizes that effective interventions must address all three barriers simultaneously.
Recent work has begun to demonstrate how these design principles can be applied to create user acceptable privacy risk awareness tools. For example, the Imago Obscura system introduces an AI-powered image privacy copilot that aimed to address all three SPAF barriers \citet{monteiro2025imago}. Their work identified five design requirements for privacy risk tools: enabling expressive articulation of concerns (motivation), increasing awareness of content-level risks (awareness), promoting informed decision-making (awareness), facilitating easy application of mitigation techniques (ability), and ensuring user autonomy (ability).

While prior work has established principles for designing privacy risk awareness tools, from privacy notices to behavioral nudges, these approaches largely focus on raising general awareness and guiding behavior in broad contexts. Recent advances in language model reasoning now open the possibility of providing users with quantifiable population risk estimates (PREs): statistical measures indicating how many people share personally identifiable information, and how uniquely identifiable a user might be as a consequence \cite{zheng2025probabilistic}. By making abstract risks more concrete, PREs could complement existing risk awareness tools and support more informed privacy decisions. Little research has examined how to effectively communicate quantified privacy risks to support user decision-making in real-time disclosure scenarios. Our work addresses this gap by investigating how different presentations of PREs affect users' privacy decision-making processes, building on the SPAF framework's emphasis on addressing awareness, motivation, and ability barriers while incorporating design recommendations from privacy notice research and behavioral nudging literature.

\section{Methodology}
We conducted a design elicitation study with the goal of understanding how people interpret, reason about, and imagine acting upon different presentations of PREs. Guided by recommended practices in speculative design for privacy \cite{wong2019bringing}, we leveraged comic-boarding alongside reflective writing to solicit these insights through surveys, following the example of prior work \cite{wu2025design, jin2022exploring}. We first envisioned 5 design variations for PREs inspired by prior literature on usable security and privacy. PRE designs were embedded within 4 different narrative vignettes depicting various threats that users might be concerned about when posting anonymously online \cite{kang2013people}, across a total of 20 comic-boards. We presented these comic-boards in an online survey deployed via Qualtrics, with 44 Reddit users recruited from Prolific (an on-demand participant recruitment platform). Each participant reviewed three randomly selected comic-boards, seeing three out of the total five PRE designs we envisioned. We then probed them with creative writing prompts to elicit participants’ perceptions, mental models, and imagined interactions with these early-stage design concepts, in order to understand what kinds of representations they found interpretable, motivating, or actionable. We synthesized the resulting themes into a set of design recommendations presented in the discussion.

\begin{table*}[!ht]
    \centering
    \caption{\parbox{\textwidth}{Population Risk Estimates alongside the respective design challenges that they try to address}}
    \label{PRETablexDesign} 
    
    \begin{tabular}{p{3.4cm}p{3.5cm}p{3cm}p{5cm}}

    \toprule
    \textbf{Design challenge} & \textbf{Prior literature} & \textbf{PRE Design} & \textbf{Description} \\
    \bottomrule

    \midrule
    Visualizing risk probability
    & Data-driven password meters from \citet{ur2017design}.
    & Re-identifiability Meter (Figure \ref{fig: Examples of Population Risk Estimate Designs}, Design \#2)
    & Show how many other people the disclosures apply to. \\

    \midrule
    Aligning imagined and real audiences when sharing online
    & Audience blindness of online posts, \citet{wang2011regretted, wang2014field}.
    & Threat-specific risk (Figure \ref{fig: Examples of Population Risk Estimate Designs}, Design \#4)
    & Display specific threat models \textit{(e.g., friends/family, organizational threat, ambiguous others, etc.)} \\

   \midrule
    Understanding impact of individual disclosures on re-identifiability
    & Privacy risk awareness in online community posts, \citet{krsek2025measuring}.
    & Risk by disclosure (Figure \ref{fig: Examples of Population Risk Estimate Designs}, Design \#5)
    & Quantify impact of individual disclosures on overall threat re-identification \\
    
    \midrule
    Interpreting PRE risk score
    & Numeracy influences risk comprehension, \citet{reyna2009numeracy}.
    & Simplified risk level (Figure \ref{fig: Examples of Population Risk Estimate Designs}, Design \#3)
    & Distill risk estimates into interpreted scores (low, medium, high) \\
    \bottomrule
    
    \end{tabular}
\end{table*}

\subsection{Design Approach}
Our goal was to explore how different presentations of PREs might impact users' perceived awareness of disclosure risks, motivation to address those risks, and ability to address those risks. In so doing, we also hoped to elicit potential comprehension barriers that users may face when making sense of these presentations of risk. 

\begin{figure*}[htp]
    \centering
    \includegraphics[width=16cm]{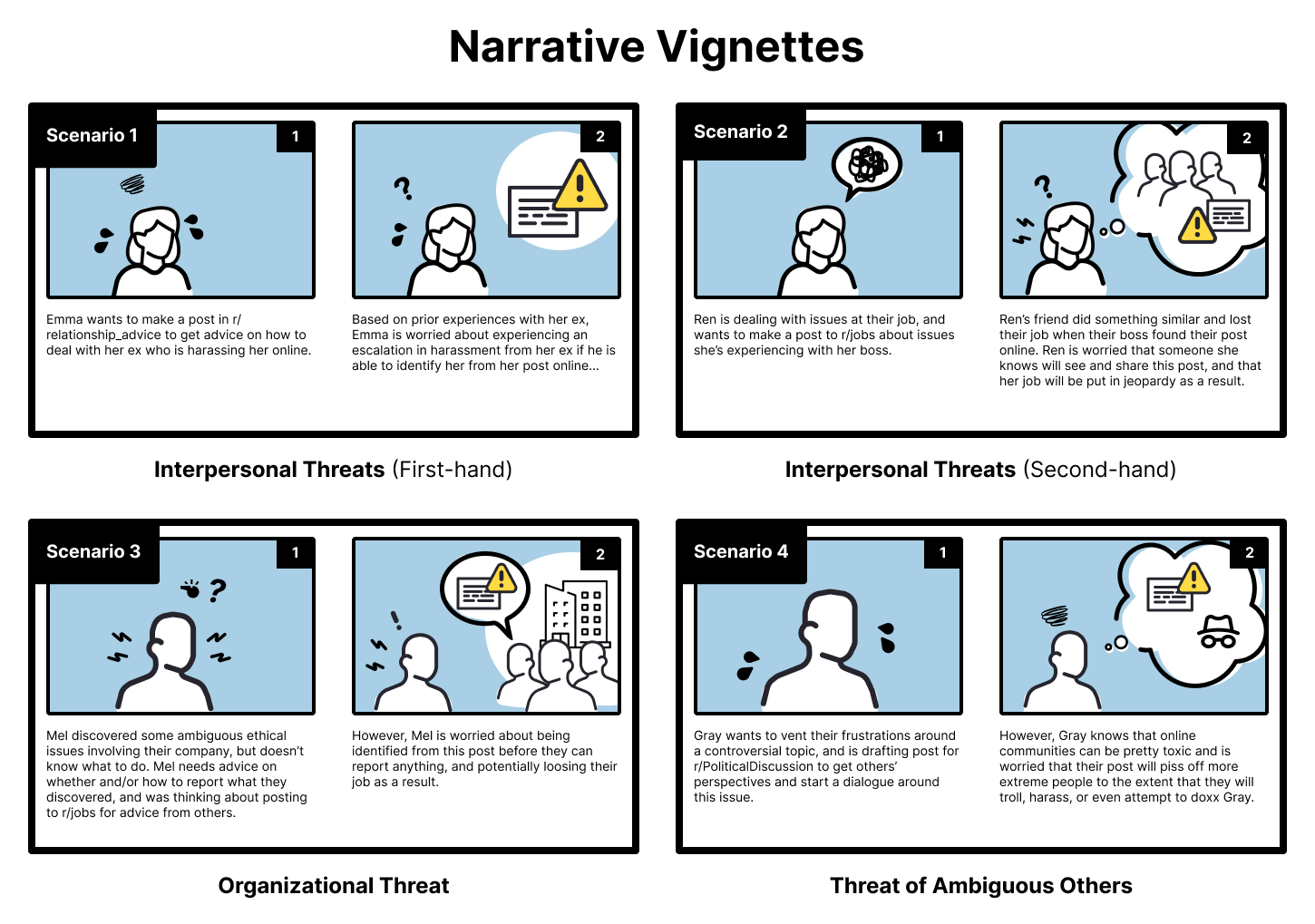}
    \caption{This image depicts each of the narrative vignette scenarios we explored in the comic-boards, from interpersonal threats to the threat of ambiguous others.}
    \Description{A depiction of each of the narrative vignette scenarios we explored in the comic-boards. In the top left, we showcase scenario 1 which featured a character who was concerned over an interpersonal threat (based on their firsthand experience with that person). Scenario 2 in the top left also depicts a character concerned about interpersonal threats, though this time based on second-hand experience (hearing about others who were re-identified and experienced harm as a result). In scenario 3 on the bottom left we have a character with concerns over an organizational threat (their company) finding their post, and then in the bottom right scenario 4 depicts a character being concerned over the threat of ambiguous others (e.g., other Reddit users, malicious actors, etc.).}
    \label{fig: Examples of Scenario Types}
\end{figure*}

\begin{figure*}[htp]
    \centering
    \includegraphics[width=\linewidth]{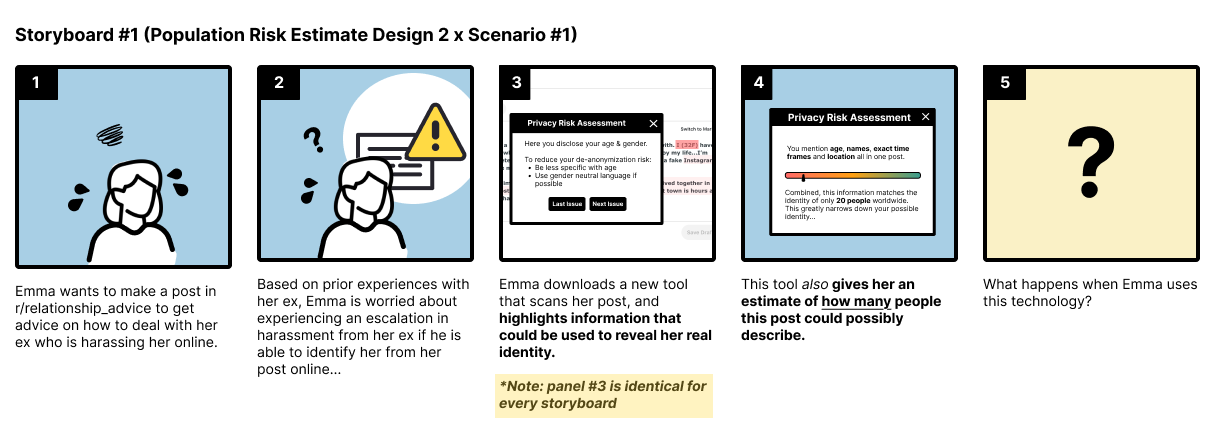}
    \caption{We use an adapted version of the comic-boarding method \cite{moraveji2007comicboarding, kuo2023understanding, jin2022exploring, wu2025design} to understand Reddit users perspectives on various population risk estimate designs (this image depicts the risk meter design inspired by Ur et. al's data-driven password meter from usable security \& privacy literature) \cite{ur2017design}. Users were randomly presented 3 out of 5 different population risk estimate designs, matched to 3 out of 4 different risk scenarios in order to elicit specific feedback and reactions around the design and deployment of this risk awareness technology.}
    \Description{An image depicting a comic-board consisting of 5 panels, representative of the comic-boards that participants saw during the study. The first two panels set the stage for the narrative vignette, describing both the main character of the comic-board and their primary concerns around privacy. The third panel introduces the concept of the privacy risk assessment tool that this character is about to interact with (panel 3 was the same for every comic-board). Then in the 4th panel, the population risk estimate design (PRE) was presented alongside a description of it's basic function. Finally, the 5th panel featured a yellow background with a big question mark in the center, and the caption below offered participants the following reflective prompt: ``What happens when [character name] uses this technology?''.}
  \label{fig:fullcomicboard}
\end{figure*}

\paragraph{Comic-Boarding as an Approach for Abstract Technologies} We employed the comic-boarding method to explore end-user perspectives on PREs. As a technique, comic-boarding involves the creation of comic-strip style comic-boards (Fig. \ref{fig:fullcomicboard} and \ref{fig: Examples of Scenario Types}) that are only partially completed, with the intention of facilitating brainstorming and eliciting design insights from participants around incomplete panels. While comic-boarding has a rich history of use in HCI more broadly \cite{kuo2023understanding, hiniker2017co, moraveji2007comicboarding}, this approach has only recently been gaining traction in usable S\&P \cite{wu2025design, jin2022exploring, holten2021work, wong2017eliciting}. A common challenge in soliciting feedback on emergent technology (particularly in the domain of S\&P) is that it is difficult for users to speculate on abstract technologies that they have never experienced before; the hypothetical seed comic-boarding affords gives participants the ability for broader ideation absent the burden of having to imagine specific implementation details, allowing them to be more imaginative with how different scenarios might impact others. Moreover, users may feel hesitant to test new disclosure privacy technologies on their own disclosures that they would prefer to keep private. Comic-boarding helps address these common challenges by situating these technologies in concrete and relatable narrative vignettes that participants can critique without divulging their own data. This technique is well suited for design-elicitation because it provides enough narrative structure to anchor participants’ interpretations, while still giving them space to envision how PREs might matter, what they would pay attention to, and how they imagine the technology being used in real-world disclosure decisions. Furthermore, it has been successful in remotely deployed formats as well (e.g., surveys) offering a good balance of depth and scalability \cite{wu2025design, jin2022exploring}. For this reason, we adapted a survey-based approach to explore more design variants with a larger pool of participants, and the asynchronous nature of the study design allowed participants to engage in the creative writing exercises of our study at their own pace, minimizing the social pressures that can emerge in in-person think-alouds.

\paragraph{Comic-board Narrative Vignettes}
We drafted four online posting scenarios within which our PRE designs were situated. Each of these four scenarios revolved around one character who wanted to make a post on Reddit to seek out support (either resources, information, or advice for navigating a situation) but who has concerns over being re-identified. We intentionally varied scenarios by the type of personal threat models characters were concerned about, and developed four scenarios inspired by prior literature exploring the key factors that motivate users to seek out anonymity online \cite{kang2013people}: risk of exposure from known others (based on first-hand experiences), risk of exposure from known others (concerns raised via second-hand concerns), risk of exposure to an organizational threat, and finally the threat of ambiguous (potentially malicious) others (see Fig. \ref{fig: Examples of Scenario Types}). For example, Scenario \#1 depicts an example of an interpersonal threat wherein the main character, Emma, wants to make a post to r/relationship advice in order to get guidance for how to deal with an abusive ex-partner who is harassing her online, but has concerns over this harassment escalating if she is re-identified by her ex from her post. By focusing on the key concerns of those who seek out anonymity online, we aimed to explore the concerns of users who would be most impacted by the use of PREs. These varied narrative contexts allowed us to examine how participants reasoned about PREs across different threat models while still keeping the overall task consistent.

\paragraph{Population Risk Estimate Designs} 
To ground our approach, we framed our designs around a specific method for quantified privacy risks based on sensitive disclosures, originally intended for measuring population-based privacy risks of a dataset -- k-anonymity \cite{sweeney2002k}. We chose this method to ground our designs as recent work from \citet{zheng2025probabilistic} has explored adapting this method to provide tailored privacy risk estimates. We decided to explore different designs for quantifying and re-interpreting this information, in order to explore whether certain presentations of PREs stuck with our participants, and explore how much granularity is needed for them to be perceived as helpful (see Table \ref{PRETablexDesign}). For Design \#2, inspired by prior literature on data-driven visual risk communication in the context of end-user passwords \citet{ur2017design}, we visualized PREs to users along a meter or spectrum (with high risk on one end, low risk on the other). Relatedly, Design \#4 drew inspiration from the work of \citet{wang2014field} who explored framing posting options in a way that attempts to close the gap in audience-blindness of online posts, citing the misalignment between their envisioned audiences and real audiences when sharing online \cite{wang2011regretted}; our design draws the connection to re-identifiability based on specific threat models a user seeking anonymity might be concerned about (e.g., friends/family, organizational threat, ambiguous others, etc.). Design \#5 was inspired by prior work citing users desire to better understand the impact of individual disclosures on their overall re-identifiability in online posts \cite{krsek2025measuring}, and as such displays a quantified impact of individual disclosures on users overall threat of re-identification. Design \#3 was designed with the intention of exploring the level of granularity necessary for PREs to be considered useful, distilling the quantified estimates of risk into an intuitive and easy to interpret level of either ``High risk'', ``Moderate Risk'', or ``Low risk''. Finally, Design \#1 serves as a control, free of any re-interpretation and allowing us to examine whether it is even necessary to re-interpret PREs to make them helpful to users. 

PREs were embedded into comic-boards with each of the four narrative vignettes, leading to a total of 20 variations, of which participants randomly saw 3. Depictions of all 20 comic-boards can be found in Section E of the Appendix. Comic-boards first introduced the scene, and then introduced the PRE design, followed finally by an invitation to complete the comic-board. For example, Fig. \ref{fig:fullcomicboard} follows the story of Emma, who wants to make a post to r/relationship\_advice in order to get guidance from others for how to deal with an abusive ex-partner who is harassing her online. Emma however, has seen authors of similar posts in the past be re-identified after sharing their stories, and has fears that her ex-partners harassment will escalate if he discovers she made this post. The fourth and fifth panels of the comic-board depict Emma encountering a new privacy risk tool (PRE) that can help evaluate her online risk, and describe how the tool works along with images of it's output. The fifth panel does not depict an image, and invites the participant to answer the question, "What happens when Emma uses this technology?". We purposefully have participants fill in the last comic-board since we wanted them to imagine how PREs might impact the outcome of characters' stories. 

\begin{figure*}[htp]
    \centering
    \includegraphics[width=16cm]{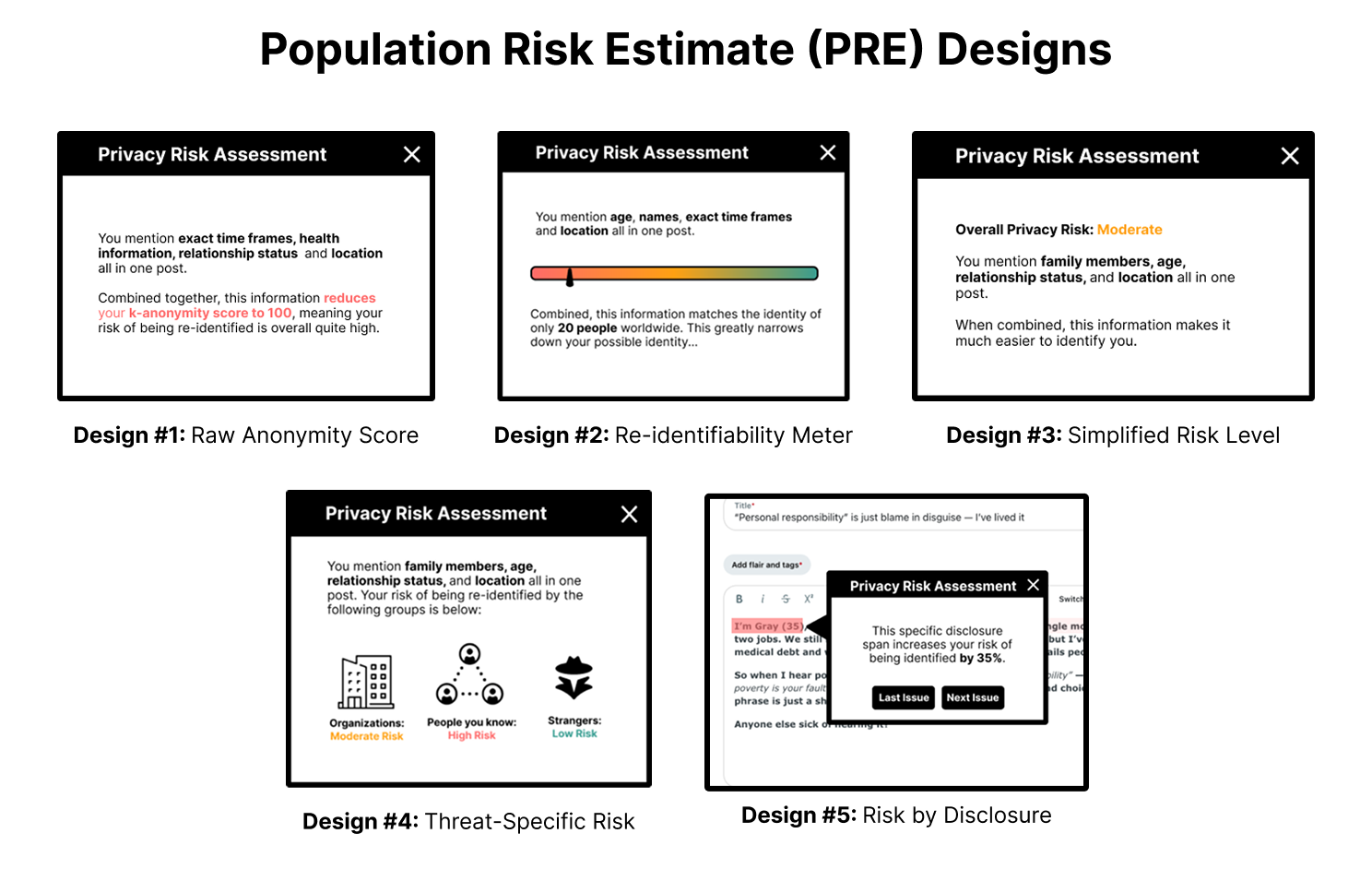}
    \caption{Image depicting our various population risk estimate designs, explored in the comic-boards.}
    \Description{Image depicting our various population risk estimate designs, explored in the comic-boards. On the top row, pictured in order from left to right are: a re-identifiability meter showcasing users re-identifiability with respect to the number of others their unique combination of disclosures could apply to (Design \#2), A raw k-anonymity score (Design \#1), and a simplified risk level where risk is communicated as being "moderate" (Design \#3.) On the bottom row, from left to right: a threat-specific risk design that showcases ris relative to different threat models a user might be concerned about such as organizational threats, people you know, or strangers (Design \#4), and finally a design showing the riskiness of individual disclosure spans (Design \#5).}
    \label{fig: Examples of Population Risk Estimate Designs}
\end{figure*}

\subsection{Survey Design}
\paragraph{Study Preface}
There were four main components to our survey. The surveys began with detailed instructions on the context of the study, explaining that participants would review three comic-boards related to a fictional world and would be asked to write a reflection on the technologies they encountered in this fictional world. We also explicitly stated that there was no right or wrong way to write their reflections, and that participants could be as creative as they like. Before allowing them to move onto the next section, we asked participants to confirm their understanding of our instructions (see Appendix, Section \ref{appen:SurveyInstructions}). 

\paragraph{Comic-Board Evaluations}
Next, participants were prompted to read and provide written reflections on three comic-boards, one at a time. To avoid cognitively overwhelming participants with tasks and maintain engagement, we followed the approach of prior work \cite{jin2022exploring} and only had participants reflect on three randomly selected comic-boards. We designed the survey logic to ensure that participants were balanced across scenario and PRE design combinations, and to ensure that each participant only saw each scenario and PRE design once. For each comic-board reflection, we asked participants three questions related to our research questions. First, to evaluate the impact of PREs on awareness they were asked to reflect on what they thought about the character's risk of re-identification after looking at the PREs (RQ1). To identify potential obstacles to action, we also asked participants about any challenges the character encountered when trying to understand the PREs (RQ3). Finally, to examine how participants envisioned PREs might contribute to successfully avoiding re-identification threats, we asked participants to write a short story describing what happens to the character in the comic-board after they see the PRE, referencing the empty panel in the comic-board (RQ2, RQ3). We intentionally placed the comic-board completion task after the aforementioned reflection questions to prime participants to include more detail in their final stories. To assess how well the proposed PRE designs could address the needs of the character, we asked participants whether they felt this design provided helpful information to the character in the comic-board, and whether the character's concerns were addressed by the tool. We had participants rate the degree to which they agreed or disagreed with these statements on a 5-point Likert scale, and elaborate on their reasoning in one open-ended question (Appendix, Section A.5). 

\paragraph{Validation Tactics}
 To validate whether the observed character's needs were aligned with the participants' needs, for each comic-board after the reflective writing exercise we asked participants to rate how much they identified with the concern. We also probed participants on the perceived riskiness of the situation characters faced in each comic-board to validate that the scenarios were perceived as equally risky. We measured both of these constructs by asking participants to rate their level of agreement on a 5-point Likert scale (Appendix, Section A.5). Finally, to ensure participants were attending to the details of the designs they viewed, after reflecting on all three comic-boards, participants were finally asked to select each of the three population risk-estimate designs that they saw --- out of all five --- and subsequently asked to rank those PRE designs in order of ``most preferred'' to ``least preferred''. Having participants select the comic-boards they saw served to confirm that participants had attended to the task and could reliably recall which designs they encountered. They were also probed on their rationale for this ranking in a single open-response question immediately afterwards. A copy of the survey can be found in the Appendix (Section A.2-A.6). 

\subsection{Recruitment}
Participants were recruited through Prolific, and were screened to ensure they were 18 years of age or older, resided in the U.S., and had an active Reddit account at the time of the study (Appendix, Section \ref{appen:Screener}). From {\Npre} participants who were screened, {\NQApre} participants were invited to proceed with the remainder of the survey. Of those {\NQApre} participants, three participants responses were removed from the final data analysis for quality assurance, as these participants failed to provide substantive answers to the open-response prompts. In total we analyzed the responses of {\N} participants. Given our sample size was reflective of those in similar prior work \cite{wu2025designx, wood2017they}, and that subsequent rounds of qualitative analysis of the three core open-ended reflections on the comic-board no longer yielded new data or themes \cite{fusch2015we} indicating thematic saturation, we did not run additional recruitment attempts. The survey took an average of 40 minutes to complete, for which participants were compensated 7 USD for their participation (at a base rate of \$10 per hour, which we rounded up). The collected demographic data is displayed in Section B of the Appendix.

\subsection{Quantitative Analysis}
Since our primary goal was to elicit participants’ reasoning about PRE designs, we used quantitative analysis mainly to check that the narrative vignettes provided a sufficiently comparable context for interpretation. Specifically, we examined whether scenarios differed meaningfully in perceived risk or relatability so we could contextualize participants’ reflections with confidence that scenario effects were not driving interpretation.

To validate that participants felt the concern described in the scenario was equally relatable and across each of the four narratives randomly presented to participants (and rule out the influence of certain narrative aspects on PRE design impressions), and that they perceived each scenario as being similar with respect to the risk posed we evaluated the responses to the aforementioned scales (see Appendix, Section A.5) across scenarios. Because participants rated only three of the four scenarios, creating a partially repeated-measures dataset, we fit linear mixed-effects models with random intercepts for participants to account for within-participant dependencies. As a robustness check we also conducted a MANOVA pairwise comparison (adjusted with Pillai’s trace to account for uneven cell sizes or small sample sizes). Both analyses yielded consistent results: there was no statistically significant impact on the perceived riskiness or the relatability of different scenarios (MANOVA: (p=0.21); LMM: all effects non-significant (see Appendix Section C.2, Tables 5, 7, \&8) A two-way MANOVA suggested a slight interaction effect between PRE design and scenario on scale rankings (p=0.03) (Appendix Section C.2, Table 6), though no significant effects were observed for rankings of the PRE designs themselves (Appendix Section C.3, Table 9) (p=0.32). To account for the partially within-subjects design, we ran linear mixed-effects models including a random intercept for each participant. These models account for correlations in responses within participants and showed no statistically significant main effects or interactions, suggesting that the small interaction observed in the two-way MANOVA does not reflect a robust effect (see Appendix Section C.3, Tables 10 \& 11). While absence of evidence is not evidence of absence, at the very least these findings suggest that there was not a clearly discernible impact of narrative vignette on PRE design impressions, if any such impact existed. 
Assessments of mean inter-item correlation of the scale items assessing the perceived risk level of the scenario, the relatability of the scenario (0.226, 0.237) (Appendix Section C.3, Table 12) were found to fit within the acceptable range (0.15-0.50) of inter-item correlation, showing that the scale items correlated well enough that they measure the same concept but were not so similar so as to be repetitive. We did not use Cronbach’s Alpha in assessing this, as with shorter scales such as ours (fewer than ten items) it is common to find quite low Cronbach values (e.g., 5) \cite{Pallant_2013}.

\subsection{Qualitative Analysis}
Our comic-boards were self-referential and inherently related, so following prior work \cite{wu2025design, kuo2023understanding, hiniker2017co} we intentionally analyzed participants’ responses across comic-boards and questions. While we did not actively screen for the use of large language models chatbots (LLMCs) by participants since there are no foolproof mechanisms for detecting AI-written content, we did deploy two prevention techniques described in literature in order to dissuade participants from using AI-written content: first restricting copy-paste into the text box, and second explicitly asking participants to commit to not using LLMCs while also clarifying that uncertain answers were acceptable to respond with \cite{traylor2025threat} (see Section \ref{appen:AIPledge} of the Appendix for a verbatim record of these techniques). These approaches have been found in prior literature to cut suspected LLMC usage by 50\% \cite{veselovsky2025prevalence}. To ensure participants were paying attention, we also had them select which three of the five total PRE designs to which they were exposed, and checked for correctness. No participants were removed from the study based on this attention check.

To analyze all 132 PRE reflections from participants (44 participants * 3 reflections per participant), we used an inductive coding approach to open-ended responses from participants. To reduce the potential for bias towards specific PRE designs to emerge in the qualitative analysis, responses were blind coded so that coders were not aware of the intervention PRE group. This analysis was conducted by two researchers who collaborated in open-coding for variations in responses revolving around the nature of the RQs that our open-response prompts were concerned with to come up with 80 initial codes, containing a mixture of general reactions to the PREs with respect to awareness of, motivation to, and perceived ability to address potential self-disclosure risks. 

The two researchers continued to discuss disagreements and consolidate codes, ultimately settling on a codebook of 18 higher-order themes and 65 sub-themes (Appendix, Section D.1). For a random subset of the study sessions (24 reflections) which were coded independently by two researchers, we achieved a Cohen’s Kappa of 0.96 — an inter-rater reliability value generally considered to represent high agreement. We then applied this codebook to the full dataset, splitting the remaining 108 responses between the two researchers. We present these themes in the following section.

\section{Findings}
The reflections we reviewed covered a wide range of perspectives on how PREs might impact users: from empowering characters to seek help without fear online, to frightening characters to the extent that they no longer felt safe posting. Overall, our findings suggest that while PREs can indeed improve risk awareness, presenting these estimates to users requires striking a delicate balance: for some users, too much emphasis on risk alone can lead to avoidance of posting entirely.
To distinguish between our participants and their reflections on the outcomes of those depicted in the storyboard, \textbf{throughout the findings, we use the term \textit{``participants''} to refer to the individuals in our study who reflected on the comic-boards, and \textit{``characters''} to refer to the fictional individuals depicted in those boards whose situations participants were asked to interpret.}


\subsection{RQ1: How PREs Influence Awareness of Risks Associated with Self-Disclosure}
\subsubsection{PREs effectively raised participants' awareness of disclosure risks. }
Across 132 participant narratives, the majority of participants (98/132, 74.24\%) (Appendix Section D.3, Table \ref{Table:PS}) described characters experiencing shifts in their risk perception after viewing the tool's output. Many responses (39/98) (Appendix Section D.3, Table \ref{Table:PS}) contained broad reflections about characters becoming generally more aware of their risk level. However, some participants went into greater detail, describing increased awareness about what could be inferred from the information disclosed in the post (65/98) and heightened sensitivity to the audience who would see the character's post (8/98)(Appendix Section D.3, Table \ref{Table:PS}). These patterns suggest that PREs reliably surfaced both general and specific facets of disclosure risk that characters might otherwise overlook.

\subsubsection{Increased awareness of risks leads to increased anxiety.}
While PREs successfully heightened awareness, these shifts were predominantly accompanied by negative emotional reactions (63/98) (Appendix Section D.3, Table \ref{Table: ERxPS}). Participants wrote about characters experiencing heightened feelings of anxiety, worry, and distress upon seeing their risk estimates. Some participants even described characters contemplating more severe forms of self-censorship, with characters questioning whether posting was worthwhile given their risk of re-identification and potential repercussions (11/63) (Appendix Table \ref{Table:ERxUA}).

On the other hand, a small number of participants (6/132) (Appendix Section D.3, Table \ref{codebook}) described characters feeling a sense of relief upon seeing PRE as they felt empowered to address the identified privacy risks. However, at least two of these participants seemed to have misinterpreted the meaning of the risk estimate, mistakenly believing the result indicated the characters were ``less identifiable'' when the opposite was true, such as in the following excerpt wherein P14 mistakenly understood the PRE (of $k=20$) not as being 1 of 20 people, but as a 1 out of 20 chance of being re-identified: \textit{``Mel is likely feeling a significant sense of relief and a boost in confidence regarding their anonymity. The panel states, `Combined, this information matches the identity of only 1 in every 20 people worldwide...'...Seeing that their combination of age, name, exact time frame, and location details only uniquely identifies them within a relatively small fraction of the global population would likely alleviate their initial worries about being identified by their workplace''}(P14).

For some participants, PREs led to extrapolated descriptions of perceived threats (11/132) and repercussions (10/132) that went beyond those defined in the original comic-board narratives (Appendix Section D.3, Table 14). For example, when asked to envision what happened to a character who sought out anonymity online to discuss politics, P29 envisioned that character having concerns over the downstream harms of their identity being doxxed, such as their family being targeted by malicious strangers who disagreed with their political take: \textit{``Gray reads the privacy tool after assessing his draft Reddit post. He sighs wondering if the potential risks to himself and his family are worth the political post if he is identified, given the current political polarization that exists within the community.''} We only noted 1 reflection from participants where these two groups overlapped. Overall, these reflections indicate that increases in risk awareness were often mentally taxing, frequently amplifying anxiety and, for some characters, pushing them toward more extreme forms of self-censorship.

\subsubsection{Misaligned conclusions of risk led to interpretability challenges and skepticism.}
Participants varied in how they envisioned characters would respond to the tool's risk assessments. While the majority of participants described characters accepting PREs at face value (100/132) (Appendix Section D.3, Table \ref{Table:RP}), approximately 24.24\% of responses (32/132) showed characters exhibiting skepticism toward the tool's output. We defined skepticism as characters drawing on their own perspective to reach conclusions about the risk level rather than accepting the tool's assessment. P17, for example, described a character's reaction: \textit{``Though the tool only assigned a 'moderate' privacy risk, this is probably too much of a risk—more than she should be willing to attempt.''}

While this skepticism was sometimes linked to characters' personal comfort with perceived risk levels, it commonly emerged from challenges related to interpreting the tool's output. Issues of transparency appeared in almost a third of the skeptical scenarios (10/32) (Appendix Section D.3, Table \ref{Table:RPxPSxOIxTrans}), as participants described characters struggling to interpret PREs. Participants wrote about characters expressing uncertainty about whether they could trust the tool or not, and whether its evaluation missed information that might impact their threat of re-identification. Challenges with interpretability also influenced reactions to tool outputs (6/32), with participants describing character concerns about how the tool makes its estimations, and the sources of its data (Appendix Table \ref{Table:RPxPSxOIxTrans}). 

Interestingly, whether participants described characters as accepting the tool's risk assessment or forming their own judgment did not predict participants' overall impression of the tool. Among the 32 participants who described skeptical characters, reactions to the tool were equally distributed between positive, negative, and mixed impressions. This finding suggests that even when participants envisioned characters questioning specific risk scores, they still recognized the tool's broader value. Characters who approached the tool's output with skepticism still benefited from its ability to draw attention to risks they might not have previously considered (11/32)[``awareness of risk'' code (Appendix Section D.3, Table \ref{Table:RPxPSxOIxTrans})] and to highlight the ``inferrability'' of information disclosed in their posts (16/32)[``information inferrability'' code, (Appendix Section D.3, Table \ref{Table:RPxPSxOIxTrans})]. Generally, skepticism did not necessarily undermine the perceived value of PREs. Instead, it highlighted how interpretability and transparency shape whether characters integrate risk estimates into their own judgments or treat them as one input among many.

\subsection{RQ2: How PREs Impact Perceived Motivation to Address Self-Disclosure Risk}
PREs generally motivated risk mitigation behaviors, largely to reduce harm to the characters themselves (first-hand harms) or to their close connections (second-hand harms). However, we observed both adaptive and maladaptive motivational responses. Adaptive motivational responses were represented by characters seeking to minimize disclosure risk while still seeking support online. Maladaptive motivational responses were less common, but still evident—often represented by characters self-censoring altogether after feeling overwhelmed or disempowered by the risk estimates.

\subsubsection{Adaptive motivational responses}
The majority of participants described characters experiencing feelings of vulnerability and concerns about potential harm to themselves or their close connections as motivating moderate self-censorship efforts, such as editing risks identified by the tool (87/132) (Appendix Section D.4, Table \ref{codebook}). This represents the most common adaptive response, where participants envisioned characters maintaining their ability to seek support through self-disclosure while taking protective measures to mitigate risks.

Participants who did not describe feelings of overwhelm often reported that PREs motivated ``goal-posting'' behavior—iterative changes to posts aimed at achieving the lowest possible risk estimate from the tool. This motivation was almost exclusively coupled with participants grounding their risk perception directly in the tool's output rather than relying on their own judgment about privacy risks (13/16) (Appendix Section D.4, Table \ref{Table: MotivxRP}).

While participants described characters often feeling empowered by this process (9/16), the experience of risk reduction still frequently created friction as characters struggled to maintain post utility while reducing identified risks (9/16) (Appendix Section D.4, Table \ref{Table: UAxMotivation}). Collectively, these patterns portray PREs as a catalyst for more cautious self-disclosure, with characters actively negotiating between protection and support-seeking rather than simply withdrawing.

\subsubsection{Maladaptive motivational responses}
A subset of participants (25/132) (Appendix Section D.4, Table 20) described how risk awareness motivated more extreme forms of self-censorship, including decisions not to post at all, delete existing posts, or leave the Reddit platform entirely. The characters that these participants described were thus unable to reap the benefits of self-disclosure they originally sought. While acts of extreme self-censorship were seen across various motivations, this reaction was the most common response in reflections where characters were motivated by a sense of privacy fatigue. This sense of fatigue stemmed from overwhelm upon seeing the risks and difficulties in editing posts to reduce risk while maintaining their communicative value, leading them to question whether posting was worth the effort as reflected in the following passage from P35, \textit{``Emma stares at the privacy assessment, her heart sinking as she realizes how much identifiable information she included in her post. The tool has confirmed her worst fear — that her ex could easily connect the dots and recognize her. Feeling exposed, Emma deletes the draft and shuts her laptop, unable to shake the sense of vulnerability. She wonders if she can ever safely seek advice online without risking her anonymity. The fear of retaliation lingers, making her second-guess every word she might share in the future.''}. It was common for participants describing characters motivated privacy fatigue to experience issues with the explainability of the PREs, needing more guidance for how to de-risk their posts (Appendix Section D.4, Table \ref{Table: UXIssuesxMotivation}), and an overarching sense of dis-empowerment in the absent this guidance (Appendix Section D.4, Table \ref{Table: UAxMotivation}). The same was true of participants who described characters motivation as stemming from a sense of vulnerability and desire to mitigate harms to themselves and peers. Extreme self-censorship among these reflections were also often accompanied by usability issues related to the PREs (e.g., explainability, interpretability, and transparency), and frustration in attempts to reduce privacy risks in their post, though feelings of dis-empowerment were lower among these reflections (Appendix Section D.4, Table \ref{Table: UAxMotivation}). These findings highlight the critical tension that while PREs can effectively motivate risk-mitigating behavior, they can also lead to counterproductive outcomes when users lack adequate support for interpreting and acting on the information provided. Therefore, PREs must balance risk awareness with actionable guidance to avoid overwhelming users and inadvertently preventing beneficial self-disclosure.

\subsection{RQ3: How PREs Impact Perceived Ability to Address Risks} Ultimately we found that the majority of reflections ended with characters successfully evading re-identification (79/132); a subset of these reflections described evading this risk as a result of refraining from posting entirely (18/72). A small number of reflections ended with participants being re-identified (11/132), but most of the remaining reflections either neglected to mention this (29/132) or described uncertainty over what might happen (describing both outcomes as equally plausible)(10/132). In the following sections, we further break down key themes that help explain how participants envisioned characters successfully and unsuccessfully balancing risk mitigation with seeking support, and the impact of PREs on these outcomes. 

\paragraph{When Participants Envisioned Characters' Success in Evading Re-identification (79/132)} The majority reflections described characters as successfully avoiding re-identification after making their post online (61/79). Of these 61 reflections, the vast majority of them described characters de-risking their posts with the aid of the PREs (55/61), with these participants describing how having these estimates empowered characters to act on potential threats and encouraged a sense of security that enabled them to seek the support they needed, as reflected by P44, \textit{``...Encouraged, he connects with a whistleblower support group and safely reports the issue. Months later, Mel looks back, grateful the tool gave him the confidence to speak up without putting himself at risk. He reflects on that moment of hesitation, when he almost didn’t post out of fear. The tool didn’t just help him reduce risk; it helped him speak out when it mattered most.''}. These findings highlight participants' perception of PREs as a means of enabling them to safely seek out support on sensitive subjects without fear of repercussions. Not everyone who successfully evaded re-identification found the PREs so easy to use, however, as a handful of participants described struggling to de-risk their post in the absence of actionable guidance. Some were able to successfully overcome this hurdle via clever means of preserving their identity (7/61): \textit{``Emma feels stuck and frustrated. The privacy tool gave her a warning about a certain phrases of her post, but it did not explain why they were risky. Unsure what to do, she deletes her original post and later she rewrites it as a fictional story with changed details. It makes her feel safer, and others still connects with it. Surprisingly, her post still resonates with others. Inspired by her feedback, the tool's developer updates the system to give clearer, more helpful for future use. The tool's team later improves it on cases like Emma's.''} (P4). Others who encountered these hurdles, however, described characters as becoming stumped and opting to engage in forms of extreme self-censorship, such as avoidance of posting altogether (15/79) or giving up after several attempts to address risks in their posts (3/79). They cited feelings of overwhelm upon seeing the PREs (11/18) and uncertainty over whether it was even safe to post (10/18), as noted in the following reflection from P35: \textit{``Emma stares at the privacy assessment, her heart sinking as she realizes how much identifiable information she included in her post. The tool has confirmed her worst fear — that her ex could easily connect the dots and recognize her. Feeling exposed, Emma deletes the draft and shuts her laptop, unable to shake the sense of vulnerability. She wonders if she can ever safely seek advice online without risking her anonymity. The fear of retaliation lingers, making her second-guess every word she might share in the future.''}. These findings suggest that while PREs can effectively empower users to seek support safely, their implementation must carefully balance risk awareness with actionable guidance to prevent paralyzing users.

\paragraph{When Participants Envisioned Character's Re-identification (N = 7, 11/132)} A small subset of reflections from participants described characters as ultimately being re-identified after making their posts. For most of these, re-identification occurred in spite of the character using the PREs to modify their posts (6/11) all vividly describing challengers around attempts to de-risk posts: \textit{``The tool's lack of steps forces Mel to make guess edits in hopes to lower her score without losing its overall message.''}(P40). Many of these reflections describe characters making their posts despite residual concerns that they may be re-identified even with the changes they made (5/11). 
Others participants envisioned characters who recognized the threat posed to them, but out of desperation for support and uncertainty over how to address those disclosure risks while meeting their original needs ultimately made the decision to post anyway, as reflected by P36: \textit{``Emma, desperate to share her story, decided to take the risk - against the advice of the tool - to make a post on Reddit. She just had to share her own side of the story and hoped the online world world would believe her. It so turned out that she was mocked and claimed to not be submissive or understanding enough which warranted the treatment she received from her ex. Just a minute percentage of people show any form of compassion or believed her. Besides, her Ex decided to come out and debunk all her narratives, even worse, calling a liar and accusing her of blackmail.''} 
Others envisioned re-identification as a direct by-product of not fully understanding how to interpret the PREs, for example: \textit{``The needle in Mel's moral compass is painfully stabbing him in the frontal lobe; he cannot be at peace until this ethical fiasco has been shared! He uses the privacy tool which kind of suggests he'll be anonymous but he disregards the fine print tool saying that it is no guarantee. The company's software engineering guru O'Brian happens to jump on the Jobs community after discussing some coding with his friend and bam! Mel has been spotted because just the other day he was talking with O'Brian and little did he know that O'Brian was loyal to the company. Mel is fired and becomes notoriety for being a nefarious whistleblower but not without creating a cataclysmic series of internal and PR events when the unethical events are revealed.''}. These cases demonstrate that even when the tool output correctly identify risks, users may still be re-identified due to inadequate guidance on how to de-risk the post, failing to account for users' desperation, or misunderstanding of the tool's purpose. Subsequently, most reflections from participants ending with characters' re-identification held negative overall impressions of the PREs (7/11), some because of the the false sense of security it imbued, and lack of clear explanations as failing characters. A handful of mixed impressions of the PREs in this group (3/11), on the other hand, acknowledged the potential helpfulness of this information, but noted that current designs left much to be desired. These mixed reactions suggest that users' trust in the tool depends not only on the tool's technical accuracy, but also on transparent communication about limitations and realistic expectations about what protection they can actually provide.

\paragraph{When the Outcome was Unclear (39/132)} Several of participants reflections described uncertainty over (10/39) the fate of characters, exploring multiple hypothetical scenarios where anything could happen within their written response. Overall, the majority of these reflections erred on the side of more positive impressions of the PREs, instead placing the responsibility for any lack-luster decisions on characters themselves (4/10); though still this communicates the idea that the PREs may not be compelling to all audiences. For example, as P9 reflects \textit{``The are only two scenarios in this case. First is she is likely to get caught by the boyfriend. Secondly she is not likely to get caught. We can not be sure of the outcome...Therefore I believe that this tool, however much we do not know the outcome, serves it's purpose in identifying whether or not she will be caught by the boyfriend.''}. Others acknowledged that while the tool was helpful in raising awareness, it's efficacy in aiding characters in avoiding re-identification was doubtful: \textit{``In the empty yellow panel there are several possibilities that might happen: if Ren makes up her mind and decides to use the tool for assistance despite all the obstacles that the prompt on the tool has mentioned, she will get the assistance she needs but not as she needed it because the tool has limitation to some issues that Ren needed to address. If Ren decides not to use the tool then she gets totally no assistance and she will continue suffering on the hands of her boss. She needs to just use the tool despite all the obstacles the tool has in order to at least some assistance rather than not using it and posting her issues and get fired by her boss.''} (P2). The remaining set of responses (29/39) did not explicitly describe characters re-identification outcomes, though most (20/29) describe attempts on behalf of the participants to utilize the PRE output as a guide for modifying their post. Echoing issues seen across the spectrum of outcomes, a core challenge participants described characters as facing was how to balancing the perceived utility of self-disclosures with the inherent risk that accompanies them (17/29), and a desire for more guidance in addressing the risks in their posts (16/29). Across these ambiguous outcomes, participants treated PREs as somewhat useful but not determinative. These findings highlight how intuitive explanations of PREs are important for enabling informed decision-making across a variety of audiences, and they need to be presented alongside actionable, tailored guidance in order to help users effectively reduce privacy risks.

\paragraph{The Need for More Scaffolding to in Aid Users in Responding to Self-Disclosure Risks}
Overall, while helpful in surfacing potential risks to most (98/132)(Appendix Section D.1, Table \ref{codebook}), as in many reflections across the outcomes outlined above, participants described characters feeling that the PREs fell short of actually helping characters address those risks (41/132) resulting in some ultimately deciding to give up in sharing online (18/132) sometimes after repeated attempts to modify the content of their post, rendering some characters unable to reap the benefits of online self-disclosure. These participants described characters feeling as though in editing their posts, they had removed too much information for the utility of posting to be retained: \textit{``She finds that she doesn't receive the community engagement that she would like. The responses that she does get don't seem to be overly helpful and are more generic and sterilized due to the lack of specific context to her situation. She has learned that if you supply an overly generic situation that you are going to receive overly generic advice and a lack of emotional attachment from the community as a matter of reciprocity.''} (P29). Among those who acknowledged their level of risk but didn't know how to address it (34/132), extreme forms of self-censorship (e.g., not posting, leaving the platform, deleting posts after making them) often emerged as a reaction to the friction participants envisioned in debating the utility of self-disclosures in communicating the nuances of their situation: \textit{``Ren finds herself overwhelmed, unsure of how to modify her post without losing its intended meaning while still protecting her anonymity.''}(P4), and often left characters feeling disempowered or voiceless, as remarked by \textit{``Emma ends up feeling alone and like no one will be able to help her out. This causes so much stress and Emma feels like she is just about to give up the will to even go about it a different type of way. Emma is just about to fall into depression, not even wanting to go to her family for help out of fear of retaliation by the ex partner.''}. This led some participants to envision characters seeking out external support from peers instead (7/34), with only a handful of those actually having their needs met in the end (5/34). Taken together, these findings suggest that while PREs are helpful interventions for risk awareness, when presented alone they don't appear to successfully help address risks – participants across all re-identification outcomes described a need for these outputs to be accompanied by additional actionable guidance for how to rewrite their posts to reduce risk without loosing meaning, as expressed by P10 \textit{``Mel finds the k anonymity score helpful but they're confused about how to raise it without removing key details''}. In other words, participants consistently framed PREs as a useful warning system that must be paired with concrete, user-friendly guidance in order to translate awareness into safe and satisfying self-disclosure.

\subsection{RQ4: How PRE Design Concepts Varied in Preference and Outcome}
Participants did not appear to prefer one PRE Design significantly more than others. Across their reactions to all PRE designs, however, we distilled key design recommendations for PREs, many of which align with and enrich prior literature on explainable AI. Specifically, across all PRE designs, participants described characters as facing hurdles with the explainability, interpretability, and transparency in the comic-boards. Below, we describe how these themes emerged from the challenges participants described characters facing; we also discuss why some designs may have been more prone to these challenges than others.

\begin{figure*}[htp]
    \centering
    \includegraphics[width=14cm]{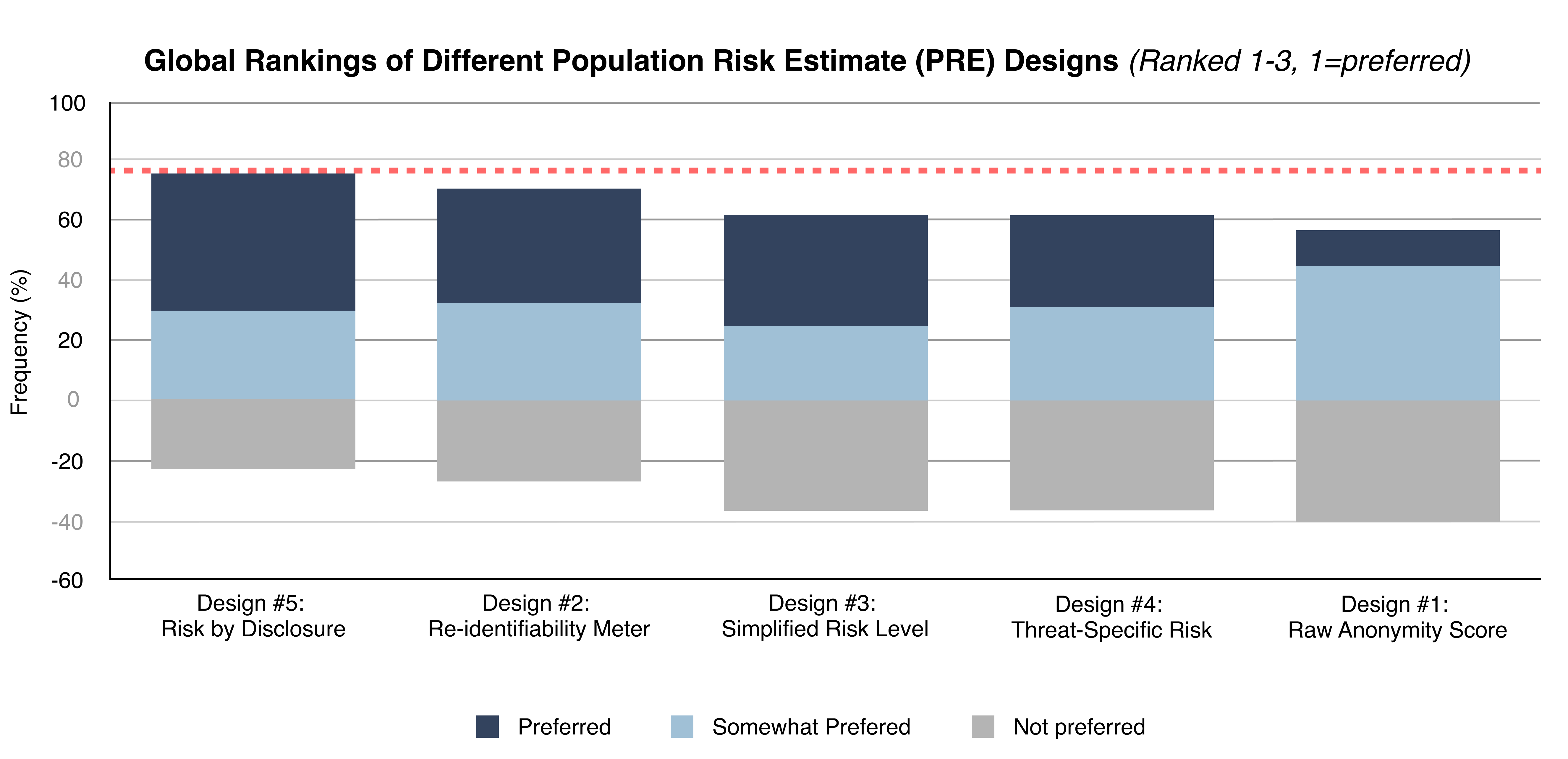}
    \caption{A stacked bar graph depicting global ranking preferences of all PRE designs. We used the Plackett-Luce method \cite{maystre2015fast, jin2022exploring, wu2025design} to merge partial rankings into a global preferred order of 5 PRE designs. PRE designs are ranked in order of preference from left to right. A higher bar indicates a more preferred PRE design. The dashed red allows for comparison across all 5 PRE designs. Design \#5 (Risk by disclosure) saw the highest popularity in rankings, closely followed by Design \#2 (Re-identifiability meter). Design \#1 (the Raw Anonymity score) was the least preferred in comparison to the other PRE designs.}
        \Description{A stacked bar graph depicting the difference in global preference rankings by PRE design on the X-axis, against the percentage of frequency with each of the PRE designs were ranked as preferred, somewhat preferred, or not preferred by participants on the Y-axis. 5 bars are shown in total. Design 5 (risk by disclosure) has the highest preference count. It is then closely followed by Design 2 (re-identifiability meter). Designs 3 (simplified risk level) and 4 (threat-specific risk) are about equal. Design 1 (raw anonymity score) is by far the least preferred of all 5 designs, ranking at the bottom.}
    \label{fig: Presentation of Self-Disclosure Spans By Model Type}
\end{figure*}

\paragraph{No single design was most preferred across all participants.}
Average ratings of PREs' perceived helpfulness, and their efficacy in addressing characters' concerns were quite high across all PRE designs (see Appendix Section C.1, Table 4), with no significant difference among them. At the end of the survey, after seeing three randomly selected PRE designs, each participant ranked the designs that they saw in the comic-boards in order of their general preference. We then used the Plackett-Luce method \cite{maystre2015fast} to merge these partial rankings into a a global order of preference for all 5 concepts (see Fig. \ref{fig: Presentation of Self-Disclosure Spans By Model Type}). Plackett-Luce is a statistical model generalized to accommodate ties of any order in the ranking. Partial rankings, in which only a subset of items are ranked in each ranking, are also accommodated in the implementation, as the method works by estimating the ``worth'' (or strength of preference) of each item based off of ranking relationships. Overall, we did not find statistically significant evidence to suggest that participants preferred any one of the designs more than the others (see Appendix Section C.1, Table 3). In terms of raw preference values, however, we found that Design 5 (Risk by Disclosure) was the most favored: 46\% of participants exposed to it ranked it as their top preference, and 23\% ranked it as their least preferred. In contrast, Design 2 (No Re-interpretation) was least favored: 41\% of participants exposed to it ranked it as least preferred, and only 12\% ranked it as most preferred. These mixed rankings suggest that no single presentation format is universally preferred, reinforcing the need for PRE designs that can flexibly accommodate diverse user goals and comfort levels.

\paragraph{Participants Had Trouble Understanding How Attackers Could Exploit Disclosures.} By far the most common challenge with PREs reported in participant's reflections was a lack of explainability (83/132), which was characterized by two core pain-points. Firstly, participants described characters feeling frustrated about the lack of context accompanying these PREs, desiring more information about why or how certain content could be used to re-identify them or how these combinations of risks might pose a greater threat to their anonymity. These issues were most salient for PRE Design 3 (Simplified Risk Level) and Design 4 (Threat-Specific Risk). Design 3 transformed raw PREs into simple ``risk'' levels --- ``low'', ``moderate'', or ``high''. While some participants praised this simplicity, others wrote about it as a source of uncertainty for characters: \textit{``I sense that she's going to spend a lot of time modifying the post, thinking it's not secure enough. Emma would really appreciate word suggestions, not hints.''} (P44). For Design 4, participants hinted at explainability issues by expressing character confusion over the threat models: \textit{``Emma likely faces challenges in interpreting the somewhat abstract risk categories like `Organizations You Know' and `People You Know' in the context of her specific Reddit post.''} (P1). 

A second common explainability issue was difficulty with interpreting the meaning of the PREs (34/132). Characters were unsure of whether certain scores indicated a high or low threat level or because the output was too technical. PRE Design 1 (Raw Anonymity Score) was most commonly implicated with this issue (12/39). Participants described the dense and data-heavy nature of the description as being a source of confusion: \textit{``Emma struggles to fully understand the implications of the privacy assessment output, as the technical language and scoring system feel unfamiliar and confusing''} (P23). Some participants even described desiring more instinctive descriptions, as P1 reflects \textit{``it doesn't explicitly explain what a k-anonymity estimate is or provide a more intuitive explanation of the level of risk associated with it. This technical term might be confusing and less impactful than a simpler, more direct warning.''}

Finally, we also noticed some participants misinterpreted PREs, as evident in their description of the risk (N=7), or it's function (N=5). Misinterpretations of the PREs output only impacted Designs 1, 2, and 5: i.e., the designs with some numeric interpretation of PRE. For Design 1, a handful of participants described interpreting the PRE (of $k=20$) not as being 1 of 20 people, but as a 1 out of 20 chance of being re-identified:
\textit{``Seeing that their combination of age, name, exact time frame, and location details only uniquely identifies them within a relatively small fraction of the global population would likely alleviate their initial worries about being identified by their workplace.''} (P14). Design 2's reflections varied in misinterpretation, for example, one participant misunderstood the value ($k=100$) as a percentage of identification risk: \textit{``She might be confused why the information she posted would 100\% tie her to her place of work''} (P37), and another interpreted it as an overall risk score: \textit{``Ren thinks that her risk of being identified is extremely high. She has a score of 100 which is the highest score.''} (P30) as opposed to communicating the number of individuals that the self-disclosures listed by the participants could apply to. Overall, these difficulties suggest that PREs must do more than display a score or label; they need to make clear how and why particular disclosures increase identifiability in order to be meaningfully actionable.

\paragraph{Participants Developed Folk Models of How PREs Were Calculated, Leading to Trust Concerns} Literature on explainable AI (XAI), describes the importance of demystifying models decision-making process on a detailed level, allowing users to trace the path taken by the model as well as how an why it reached a particular conclusion based on it's input data \cite{ribeiro2016should}. The lack of \textbf{transparency} in what data was being considered was a source of skepticism toward the PREs (N=12, 16/132), who expressed concerns over whether there might be any unrealized threats overlooked by the PRE described in the comic-board, and whether the estimate is \textit{really} capturing everything that could be used to re-identify them (such as inference-based privacy risks made by aggregating a user's posting history, or other publicly accessible information). This manifested in reflections through participants envisioning characters being re-identified from information that went overlooked by both characters themselves and the PRE in the comic-board: \textit{``Removing a city name might not be enough if her workplace has unique traits that can be easily identified. Mel debates the accuracy of the tool's `k-anonymity score.' ''} (P40). For the most part these concerns occurred with the same frequency across different PRE designs, except design 2 (Re-identifiability Meter). We suspect this may be because this design puts users' re-identifiability in direct context of a wider population (as opposed to designs that are less granular like \#3 that only vaguely characterize the threat level), prompting users to consider other factors that could distinguish them from others around them (e.g., writing style, Reddit username. etc.). 

Moreover, for AI to be widely adoptable, it must not only be transparent as emphasized in the existing literature on XAI \cite{ribeiro2016should}, but users must also be able to \textit{interpret the descriptions of these inner workings} to feel that they can trust it's outputs \cite{eslami2015always}. Our findings echo this, citing \textbf{interpretability} as a key issue across PREs and leaving participants with trust concerns over the outputs of the PREs. While for a small number of participants data driven outputs were preferred as the numbers were taken as a sign of being more trustworthy since it was relying on some vague statistics (Design 1 \& Design 2), for most this was not the case. In fact, rather this raised concerns for some participants (N=11, 17/132) about four of whom described these issues across several PRE designs – these participants expressed characters struggling to understand the inner workings of the PREs, questioning how these PREs were calculated and describing frustration over the lack of clarity with respect to the reasoning behind the PREs. The frequency at which these concerns arose were consistent across PRE designs with the exception of design 5 (Privacy Impact Per-Disclosure) which saw slightly more, perhaps because the steps needed to get to this calculation appeared more complex. Beyond just calculating the identifiability of a post author using the details in their post, this design takes the extra step of having to estimate the degree of severity of that disclosure based on how much information is there and take into consideration how much of an impact it has on the post author's overall identifiability. This understandably became a source of confusion for some participants, as seen in the following reflection from P30 where they are attempting to make sense of this calculation: 
\textit{``One obstacle is how great the odds of being identified were in the first place. For example, if the odds of being identified to begin with were 3\%, then the location would increase the odds to 4\% chance. If the odds were 60\%, then the location would increases the odds of being identified to 80\%. Another is how much each piece of info combines with others to increases the odds overall since each piece of info doesn't take place in a vacuum.''} 
Participants also described skepticism over the the source of the data, as P32 explains: \textit{``Who is gathering the data? How many people is it surveying to determine how identifiable he is? What if there is a software glitch and the data presents some unrealistic numbers and Mel's post turns out to be unique? Potential doomsday for Mel!''} To support interpretability, some participants described a desire for deeper explanation of how different combinations of categories of disclosure may compound risk, as well as more details on how the PREs work in order to encourage their confidence in the performance of such tools. 

That said, participants still found value in the PREs even while acknowledging their potential shortcomings. For example, two participants described characters sharing complaints with the creators of it so that they could use it with peace of mind in the future. For example, as P43 wrote, \textit{ ``She comes back to share her experience about this forums and warn others of possible mishaps. The discussion grows and reaches the developers of the tools prompting them to work on improving them to more secure and reliable versions. She then feels some sense of relief once the tools are worked on and corrected effectively.''} These folk models and trust concerns suggest that PREs must explicitly surface their assumptions, data sources, and limitations if they are to foster durable, well-calibrated trust rather than confusion or misplaced confidence.
\section{Discussion}

\subsection{Design Recommendations for Population Risk Estimates}
While our comic-boards were were grounded in online self-disclosure scenarios on pseudonymous platforms, participants’ reflections revealed broader insights into how AI-generated privacy risk estimates might be designed to support users across a diverse set of online contexts. In what follows, we outline four design recommendations for presenting quantified privacy risks derived from the themes surfaced in our findings. These recommendations build on the strengths of PREs --- such as raising awareness and improving users’ ability to reason about disclosures --- while also addressing unintended drawbacks. Our goal is to highlight how PREs can be designed to maximize benefits, and minimize negative externalities. Although our study included four narrative scenarios corresponding to different threat models, participants’ reflections revealed no substantial differences in how they interpreted PREs or envisioned characters responding to them. As a result, the design recommendations that follow reflect patterns that generalized across all threat types represented in our study.

\paragraph{Recommendation \#1: To Improve Users' Understanding of the Consequence of Specific Disclosures, PREs Should Be Accompanied with Explanations of How Disclosures May Be Exploited.} While the PREs were seen as helpful overall, several participants envisioned characters experiencing difficulty with understanding how attackers could exploit specific disclosures in their online posts. While the PREs identified categories of self-disclosure present in post drafts and tied them to an overall estimate of characters re-identification risk, they did not explain \textit{why} these disclosures were risky.

As a result, participants described uncertainty over the practical implications of disclosures. For example, they questioned how sharing certain combinations of information could compound risk, and uncertainty over why particular details (e.g., workplace, location, occupation) might be more identifying to one perceived threat model over another (this concern was raised several times in responses to design \#4). Absent this contextualization, participants imagined characters repeatedly guessing at what to change in their posts - goal-posting their post edits around repeated re-scans of their posts rather than confidently taking informed action. These findings indicate how PREs alone are insufficient for motivating effective risk mitigation attempts. To motivate informed action, PREs should be presented alongside explanatory feedback to users that help make privacy risks more concrete. For example, such explanations could highlight how disclosing both one's hometown and occupation could allow organizational threat models to triangulate identity, or how mentioning one's school level (e.g., high school vs. grade school) might make it easier to track their routines. By incorporating explanations of \textit{how} threat models could exploit this information, PREs could empower users to strategically reduce the risks inherent in their post without unnecessary self-censorship.

\paragraph{Recommendation \#2: To Build Trust with Users, PREs Should Be Presented with Explanations of How Estimates Were Calculated} Several participants questioned how PREs were calculated, what data they relied on, and whether they truly captured all threats that could lead to characters' re-identification (e.g., Reddit username, writing style, etc.). In the absence of transparency around the inner workings of the model creating these PREs, participants developed their own ``folk models'' of these calculations that overlooked risks or introduced software errors, undermining characters confidence in the PREs. This skepticism did not reflect rejection of the tool itself, but rather uncertainty about what exactly the model was doing.

This pattern has been documented in prior literature on end-user facing AI as well, explaining how since the operations of algorithms are often opaque, users will typically develop theories about the algorithm in order to plan or reflect on their behaviors \cite{eslami2016first, devito2018people}. Such work highlights the importance of introducing transparency into algorithms that are integrated into end-user facing systems (like PREs). Users need clarity over how a model operates, the kinds of data it draws on to make it's calculations, and any assumptions underlying it's conclusions in order to PREs to be helpful. As noted by our findings, without this visibility, users' trust in PREs could be fragile and prone to erosion. 

Equally important is interpretability, as transparency alone is insufficient if the information provided around the workings of the model is too technically complex to be digestible by users. Prior work emphasizes how the information provided on model's inner workings must also be balanced by limits of what information is practically useful to users \cite{eslami2016first}. For PREs to foster trust and provide effective decision-making support, explanations of the model's logic must be provided in ways that users can understand and apply. Many participants imagined characters struggling to reason about how the risk of compounding disclosures can end up as a simple singular numerical output, illustrating how inexplicable PREs can leave users uncertain. Accompanying PREs with interpretable explanations for estimate was made can help users make informed decisions over how much trust to place in the outputs. 


\paragraph{Recommendation \#3: To Avoid Dis-empowering Users, PREs Should Provide Suggestions for Reducing Risk While Preserving Communicative Intent}
    While PREs effectively raised privacy awareness, participants frequently described characters feeling dis-empowered when they recognized risks but lacked guidance on how to address them. This dis-empowerment manifested most clearly when characters attempted to balance post utility with risk reduction--struggling to maintain meaningful communication while reducing identified privacy risks. Participants described scenarios where characters would edit repeatedly, only to find they had removed too much information for their posts to retain communicative value. This led to posts that received only generic responses due to lack of specific context, defeating the original purpose of seeking targeted advice.
    
    The insufficient guidance had serious consequences. Some characters chose not to post at all rather than risk inadequate editing, while others developed privacy fatigue as they became overwhelmed by the complexity of balancing privacy and communication needs. The tools that were meant to empower safe disclosure instead became barriers to accessing the social support users were seeking. When PREs highlighted risky disclosures without suggesting alternative wordings, users were left to navigate complex trade-offs between privacy and communication effectiveness on their own. Thus users need concrete guidance on how to rephrase their thoughts rather than simply being told what was risky.
    
    To address these limitations, PREs should be coupled with specific, contextual guidance that helps users understand not just what information poses risks, but how to communicate their core message while mitigating those risks. This includes suggesting alternative phrasings, recommending which details are most versus least critical for their communicative goals. Without such scaffolding, PREs risk creating awareness without enabling action, potentially leading to the counterproductive outcome of preventing beneficial self-disclosure.

\paragraph{Recommendation \#4: To Avoid Misinterpretations, PREs Should Be Presented in Intuitive Natural Language}
PREs must be presented in formats that users can easily and accurately interpret to enable effective privacy decision-making. Despite the high precision of quantified risk estimates, they can be difficult for people to understand and easily lead to misinterpretations.
Indeed, we observed misinterpretations of PREs across multiple participant responses, particularly among designs that relied heavily on numeric outputs (designs \#1, \#2, \& \#5). For example, one participant misunderstood the k-anonymity value, interpreting $k=20$ as indicating a one out of 20 chance of re-identification rather than correctly understanding it as meaning the character was one of 20 similar people in the population (P14). Similarly, participants struggled with the technicality of k-anonymity, one response indicating confusion about the \textit{``dense and data-heavy information presented in the assessment''} (P23), while another described the output as \textit{``foreign and new methodology''} (P13). 

These misinterpretations and perception of k-anonymity being too technical are important to address. When users misunderstand their risk level--believing they are safer than they actually are--they may unwittingly disclose information that leaves them vulnerable to re-identification. Conversely, misinterpretations that overestimate risk could lead to unnecessary self-censorship. To avoid such consequences, PREs should employ clear, jargon-free language that communicates risk levels in terms users can easily understand.
The explainable AI literature suggests that natural language expressions are often more interpretable and preferred \cite{liu2020intuitive, wallsten1993preferences}. 
Accordingly, rather than presenting raw k-anonymity scores, effective designs should translate these technical outputs into intuitive risk descriptions. For instance, instead of ``$k=20$,'' a more interpretable presentation might state \textit{``The personal information you have shared in this post could narrow your identity down to 20 people--a moderate privacy risk.''} Such presentations maintain the precision of the underlying estimate while making the implications clear to users.

\subsection{Limitations \& Future Work}
\paragraph{Participant Recruitment \& Study Context}
While the participant demographics were evenly split on gender, our population was skewed white (68.2\%), with the majority of participants below the age of 50 (88.6\%), though these skews are fairly representative of Reddit demographics \cite{Liedke_Wang_2023}. Additionally, our sample was conducted solely with participants residing in the United States. Our participants were all recruited from Prolific, and as such it's possible that we introduced additional biases into our sample as crowd-sourced participants are accustomed to participating and volunteering in research, and on the whole tend to be more tech-savvy than the general population \cite{redmiles2019well, peer2017beyond}. 

Our survey was also quite long, averaging around 40 minutes for respondents. To mitigate the impact of survey fatigue for respondents we employed a variety of strategies, the first of which was to set expectations through clearly communicating to participants that they would be engaging in a 40-minute creative writing exercise prior to joining the study, and again at the start of the online survey. We also included visual aids throughout the survey such as progress bars to communicate participants progression through the study, leveraged hierarchy of text and visual aids to make the prompts easier to parse, and towards the end of the survey we included visual depictions of the PRE designs participants saw to serve as a memory aid when ranking them against one-another. While these tactics can reduce survey fatigue, they don't completely eliminate them \cite{jeong2023exhaustive}. Increased abandonment of surveys can be a sign of fatigue; we did not see this pattern emerge in our study. Shorter or unrelated responses can also be an indicator \cite{galesic2009effects}, so as described in the methodology we filtered out low-quality responses from participants from the date analysis (though these participants were still compensated for their efforts).

We also note that the scope of our study was limited to the context of online self-disclosure on pseudonymous or anonymous online platforms (e.g., Reddit). Because Reddit culture is characterized by candid support-seeking and detailed storytelling, participants may have imagined disclosure practices and privacy considerations that align with this platform’s norms. As a result, some of the strategies participants envisioned and the ways they interpreted PREs may reflect Reddit-specific expectations around anonymity, audience size, and conversational tone.
While the four narrative scenarios we designed reflected the most common threat models motivating anonymity-seeking behavior identified in prior work (e.g., concerns about known others, organizational risks, and ambiguous malicious actors) \cite{kang2013people}, there may be other, less common scenarios that our vignettes did not capture. Future work could explore how PREs for privacy might be received in other contexts as well, such as in the context of providing support for journalism, posting online political dissent, or other situations where users seeking anonymity may be at disproportionate risk of harm. Examining receptions to PREs in other contexts may prove helpful.

Finally, as a byproduct of attempting to acutely capture the nuance across threat models identified by prior literature in our narrative vignettes, the language around the potential harm used to describe said threats varied. While we weren't running a controlled experimental study in this work, we wanted to ensure that the scenarios were similar enough to contextualize how participants reacted to different PRE designs. As such we attempted to account for this variability by measuring the perceived riskiness of each scenario participants encountered. We found no significant difference in perceived riskiness across the different scenarios. Future work evaluating the impact of PRE tool in controlled experiments should take care to standardize the framing of harm across scenarios.

\paragraph{Design Fiction and Participant Outlook}
Though we had almost an equal number of participants express mixed feelings towards the PREs, it is possible that social desirability bias may have led some participants to over-describe positive feelings towards the PREs. Therefore, we carefully dissect and explore sub-themes on limitations and frustrations around PREs. Asking participants to compare across a handful of population risk estimate design concepts also helped to alleviate this effect.
\section{Conclusion}
In our work, using design fictions and comic-boarding, we explore five different design concepts for presenting population risk estimates (PREs) to users. Through an online survey with 44 Reddit users, our findings show that PREs can improve risk awareness and motivate informed self-disclosure. Our findings also show how PRE designs can suffer from issues of explainability, transparency, and interpretability, which if left unaddressed could dis-empower users by promoting excess self-censorship. From these findings, we distilled four key design recommendations for how PREs should be presented in order to promote risk-informed, confident self disclosure. PREs must (i) be accompanied with actionable suggestions for preserving communicative intent while reducing risk or alternative methods to seek support when the risk is too high; (ii) explain how the value of the population risk estimate was determined, with plausible ways attackers might exploit these disclosures; (iii) communicate risks in a way that promotes careful behavior without causing users to censor themselves unnecessarily; and, (iv) use clear, interpretable language and visuals that avoid technical jargon and misinterpretation. Advancements in privacy risk assessment technologies pose new challenges for the presentation of privacy notices. Current approaches to privacy-risk identification typically also only analyze a user’s content in isolation. In practice however, privacy risks frequently arise from the inferences made by aggregating information such as a user’s posting history, cross-platform footprints, and other publicly accessible information. Our design recommendations offer a starting point to reason about meaningful presentations of PREs, and can support future work exploring how to design user-facing privacy notices that draw on realistic inference and aggregation conditions. 

\begin{acks}
We thank our participants for their time and input that shaped this research. We also thank Sanjay Kairam, Daniel Amon-Kotey, Kevin Huang, Pradyumna Shome, and our anonymous reviewers for their insightful feedback on earlier drafts of our manuscript. This work was supported by the National Science Foundation (NSF) under SaTC Award No. 2316287.
\end{acks}

\bibliographystyle{ACM-Reference-Format}
\bibliography{Citations}
\clearpage
\appendix

\section{Survey Design}
    \subsection{Screener Questions}
\label{appen:Screener}
\begin{enumerate}
    \item \textbf{Please indicate your age range.} 
            \begin{enumerate}
                 \item 17 years or younger 
                 \item 18-29 years 
                 \item 30-49 years
                 \item 50-64 years 
                 \item 65 years and older
            \end{enumerate}
            
    \item \textbf{Are you currently residing in the U.S.?} 
            \begin{enumerate}
                 \item No 
                 \item Yes
            \end{enumerate}
    
    \item \textbf{Do you currently have a Reddit account?} 
            \begin{enumerate}
                 \item No
                 \item Yes
            \end{enumerate}
\end{enumerate}
 
\subsection{Survey Instructions}
\label{appen:SurveyInstructions}
\begin{enumerate}
    \item \textbf{\underline{Study Information}}
    In this study, you will be asked to read about parts of a fictional world, and write stories that take place in this fictional world. We’ll present to you various panels of images and text related to a story, and you will write what happens in the empty panels. 
    \newline \newline Each of the stories you'll see today is about a fictional character who wants to make a post on Reddit, but is concerned about preserving their anonymity. Each character will be using \textbf{a different technology} to try and address their concern– as you read the storyboards, please reflect on \textbf{whether this tool actually addresses their concerns or not, and whether they provide helpful information.}  
    \newline \newline Our goal is to \textbf{ \underline{understand your feelings}} around the technology presented in these storyboards. 
    \newline \newline There is no right or wrong way to complete the story, and you can be as creative as you like. We are most interested in your reactions to the tools described in the stories. Some names and concepts might also appear in the real world, but when responding, please assume that they exist only within the fictional parameters we will present to you. Don’t spend too long thinking about what might happen next—just write about whatever first comes to mind.
    \newline

    \item \textbf{Based on the information presented above, which of the following is true?} 
    \newline \textit{(Please re-read the information above carefully if you are not sure.)}
        \begin{enumerate}
             \item I will see panels of both images and text in this study. 
             \item I will see panels of only images in this study. 
             \item I will see panels of only text in this study.
        \end{enumerate}

    \item \textbf{Based on the information presented above, which of the following is true?} 
    \newline \textit{(Please re-read the information above carefully if you are not sure.)}
        \begin{enumerate}
             \item I can be as creative as I want when writing my free responses to the stories in this study. 
             \item I must adhere to very strict rules about what I can write during this study.
        \end{enumerate}
\end{enumerate}

\subsection{Pledge to Refrain from AI Use}
\label{appen:AIPledge}
\begin{enumerate}
    \item We ask that you also agree \textbf{to \underline{not} use AI} (e.g. ChatGPT, Claude, etc.) when answering our open response questions in the survey. The use of AI in answering open-ended responses drastically harms the quality of answers we hope to collect, since what we care about is \textbf{\underline{\textit{your}} perspective} (not that of a generative agent). If you are unsure, please just do your best to answer the question asked or explain why you weren't certain how to answer that question. 

        \begin{enumerate}
             \item I understand that AI is prohibited for this survey, and \textbf{agree not to use it.} 
             \item I \textbf{do not agree} to avoid the use of AI in my answers. 
        \end{enumerate}
\end{enumerate}

\subsection{PRE Individual Comic-Board Open Response Questions}
\label{appen:PRE Scenario Reaction Questions}
\begin{enumerate}
    \item PRE Scenario Reaction Questions
        \begin{enumerate} 
            \item Open Response Q1: What does [character name] think about their \textbf{risk of being identified} by [personal threat model] \textbf{\textit{\underline{after}}} looking at the \textbf{\textit{privacy assessment results}} of this tool \textbf{\textit{(panel \#4)}}? 
            \newline \textit{Please make sure your response is at least 3 lines long.}
            
            \item Open Response Q2: What obstacles does [character name] encounter when trying to understand the \textbf{\textit{privacy assessment results}} of \textbf{\textit{this tool (panel \#4)}}, and when attempting to address the issues surfaced by it? 
            \textit{Please make sure your response is at least 3 lines long.}

            \item Open Response Q3: \textbf{Given your responses above, what happens in the empty yellow panel.} Feel free to write as much as you like about how Emma or any other characters you come up with are impacted by \textbf{\textit{this tool (panel \#4)}}, and go as far into the future as you like.  Again, be as creative as you would like. 
            \textit{Please make sure your response is at least 5 lines long, and spend at least 5 minutes writing your story.}
        \end{enumerate}
\end{enumerate}

\subsection{Individual PRE Design \& Narrative Vignette Rankings}
\label{appen:PRE Scenario Reaction Questions}
\begin{enumerate}
    \item PRE Scenario Reaction Questions
        \begin{enumerate} 
            \item Please rate the degree to which you agree or disagree with the following statements regarding the storyboard you just read... \textit{**Responses were on a 7-point Likert scale from Strongly Disagree (1) to Strongly Agree (7).**}
    
            \begin{enumerate}
                \item I \textbf{could relate} to this character's concern (panel 2).
                \item The tool in this storyboard (panels 3-4) \textbf{addresses} this character's concerns. 
                \item The \textbf{privacy assessment results} of the tool in this storyboard (panel 4) provide helpful information to this character.
                \item I felt that the situation this character was facing is \textbf{risky}.
            \end{enumerate}
            
            \item Please elaborate on your answers to the above statements in detail. For each statement, explain why you agree/disagree.  \textit{**Open response**}
        \end{enumerate}
\end{enumerate}

\subsection{Overall PRE Design Rankings \& Rationale Questions  }
\label{appen:PRE Overall Ranking Questions}
\begin{enumerate}
    \item PRE Design Ranking
        \begin{enumerate} 
            \item Please rank the tools you saw in order of your preference for the technology displayed in them. In other words, what tool (if any) would you want to exist the most? 
            \newline \textit{**(1 = You like it the most, 3 = you like it the least)**} 
                \begin{enumerate}
                \item Rank 1: [insert tool]
                \item Rank 2: [insert tool] 
                \item Rank 3: [insert tool]
            \end{enumerate}
            
            \item Please explain the rationale behind your rankings for each of the tools you saw. What were the pros and cons of each of the tools you saw? How do they compare to one another? 
            \textit{**Open response**}
            
        \end{enumerate}
\end{enumerate}
    \clearpage

\section{Participant Demographic Summary Stratified by Gender}
    \label{Demographics}
    \begin{table}[hbt!]
{%
\begin{tabular}{@{}lllllll@{}}
\toprule
                                     &                                                    & Female (N=22)        & Male (N=22)          & Total (N=44) \\ \midrule
\multirow{5}{*}{Age}                 & 18-29                                              & \texttt{9 (40.9)}    & \texttt{4 (18.2)}    & \texttt{13}   \\
                                     & 30-49                                              & \texttt{9 (40.9)}    & \texttt{17 (77.3)}   & \texttt{26}   \\
                                     & 50-64                                              & \texttt{3 (13.6)}    & \texttt{1 (4.5)}     & \texttt{4}   \\
                                     & 65 years and older                                 & \texttt{1 (4.5)}     & \texttt{0 (0.0)}     & \texttt{1}   \\ \midrule
\multirow{2}{*}{Transgender}         & Yes                                                & \texttt{2 (9.1)}     & \texttt{3 (13.6)}    & \texttt{5}   \\
                                     & No                                                 & \texttt{20 (90.9)}   & \texttt{19 (86.4)}   & \texttt{39}  \\ \midrule
\multirow{4}{*}{Ethnicity}           & South, Southeast, or Southwest Asian               & \texttt{0 (0.0)}     & \texttt{2 (9.1)}     & \texttt{2}   \\
                                     & Black/African                                      & \texttt{6 (27.3)}    & \texttt{5 (22.7)}    & \texttt{11}   \\
                                     & Black/African, East or Central Asian               & \texttt{1 (4.5)}     & \texttt{0 (0.0)}     & \texttt{1}   \\
                                     & Caucasian                                          & \texttt{15 (68.2)}   & \texttt{15 (68.2)}   & \texttt{30}  \\ \midrule
\multirow{4}{*}{Education}           & Graduate degree                                    & \texttt{5 (22.7)}     & \texttt{5 (22.7)}   & \texttt{10}   \\
                                     & Bachelor's degree                                  & \texttt{13 (59.1)}    & \texttt{10 (45.5)}  & \texttt{23}  \\
                                     & Some college                                       & \texttt{4 (18.2)}    & \texttt{4 (18.2)}    & \texttt{8}  \\
                                     & High school degree                                 & \texttt{0 (0.0)}     & \texttt{3 (13.6)}    & \texttt{3}  \\ \midrule
\multirow{6}{*}{Employment Status}   & Full-time                                          & \texttt{13 (13.6)}     & \texttt{16 (72.7)}    & \texttt{30}  \\
                                     & Part-time                                          & \texttt{4 (18.2)}    & \texttt{1 (4.5)}    & \texttt{5}  \\
                                     & Self-employed                                      & \texttt{0 (0.0)}     & \texttt{1 (4.5)}           & \texttt{1} \\
                                     & Unemployed \& looking                              & \texttt{0 (0.0)}    & \texttt{1 (4.5)}    & \texttt{1}   \\
                                     & Unemployed \& not looking                          & \texttt{2 (9.1)}    & \texttt{0 (0.0)}    & \texttt{2}   \\
                                     & Student                                            & \texttt{3 (13.16)}  & \texttt{3 (13.16)}    & \texttt{6}   \\ \midrule 
\multirow{5}{*}{Income}              & \$100k+                                            & \texttt{7 (31.8)}           & \texttt{7 (31.8)}     & \texttt{14}   \\
                                     & \$75k-99k                                          & \texttt{7 (31.8)}     & \texttt{2 (9.1)}    & \texttt{9}   \\
                                     & \$50k-74k                                          & \texttt{2 (9.1)}    & \texttt{7 (31.8)}     & \texttt{9}   \\
                                     & \$25k-49k                                          & \texttt{6 (27.3)}    & \texttt{5 (22.7)}     & \texttt{11}   \\
                                     & \textless\$25k                                     & \texttt{0 (0.0)}    & \texttt{1 (4.5)}           & \texttt{1}   \\ \midrule
\multirow{2}{*}{Used Throwaway Account?} & Yes                                                & \texttt{7 (31.8)}     & \texttt{8 (36.4)}    & \texttt{15}   \\
                                         & No                                                 & \texttt{15 (68.2)}   & \texttt{14 (63.6)}   & \texttt{29}  \\ \midrule     
\multirow{4}{*}{Reddit Use Frequency} & More than 8 times per week                        & \texttt{8 (36.4)}    & \texttt{9 (40.9)}     & \texttt{17}  \\
                                      & 4-7 times per week                                & \texttt{ (40.9)}    & \texttt{6 (27.3}    & \texttt{15}   \\
                                      & 1-3 times per week                                & \texttt{4 (18.2)}     & \texttt{7 (31.8)}    & \texttt{11}   \\
                                      & Less than once per week                           & \texttt{1 (4.5)}     & \texttt{0 (0.0)}           & \texttt{1}   \\ \bottomrule
\end{tabular}%
}
\end{table}
    \clearpage

\section{Quantitative Analyses}
\subsection{Plackett-Luce Results of Global Preference Rankings of PRE Designs}
    \begin{table}[!htb] 
\label{PL}
        \begin{tabular}{llllll}
            \toprule
            Factor & Estimate & std error & z value & \emph{p}-value \\
            \midrule
            Design \#1  & 0.6  & 0.46 & 1.32 & 0.1868   \\
            Design \#2  & 0.00 & NA   & NA   & NA       \\
            Design \#3  & 0.25 & 0.40 & 0.64 & 0.5253   \\
            Design \#4  & 0.24 & 0.47 & 0.49 & 0.6182   \\
            Design \#5  & 0.75 & 0.41 & 1.79 & 0.0721 . \\
            Residuals & 128 \\
            \bottomrule
        \end{tabular}
            \caption{Results of a Plackett-Luce test with Design \#2 (Raw Anonymity Score) as the reference, shows no significant difference across preferences for PRE designs. Design \#5 (Risk by Disclosure) nears significance (p=0.07) 
            \\ *\emph{p}<0.05, **\emph{p}<0.01, ***\emph{p}<0.001}
\end{table}
    \begin{table}[!htb] 
\label{LikertsByDesign}
        \begin{tabular}{llllll}
            \toprule
            Factor & Concern Addressed (SD) & Helpful (SD)\\
            \midrule
            Design \#1  & 5.86 (0.96) & 5.67 (1.66) \\
            Design \#2  & 5.36 (1.39) & 6.29 (0.6)  \\
            Design \#3  & 5.62 (1.05) & 5.90 (0.98) \\
            Design \#4  & 5.38 (1.24) & 5.77 (0.99) \\
            Design \#5  & 5.50 (1.11) & 5.93 (1.05) \\
            \bottomrule
        \end{tabular}
            \caption{Average Likert Scale Rankings and standard Deviations with respect to the Likert scales on PREs perceived helpfulness, and how well PRE was able to address the concerns of characters in the comic-boards.}
\end{table}

\subsection{Results of MANOVA \& Linear Mixed Effects Model on Scenario and Relatability \& Risk Scale Rankings}
\begin{table}[!htb] 
        \begin{tabular}{llllll}
            \toprule
            Factor & DF & Pillai Approx. & F & \emph{p}-value \\
            \midrule
            Scenario  & 3 & 0.063781 & 1.4055 & 0.2128 \\
            Residuals & 128 \\
            \bottomrule
        \end{tabular}
            \caption{Results of a one-way MANOVA to explain any significant differents in narrative vignette rankings across the scale items on relatability and risk level. The results show no significant impact of scenario type across overall relatability of the scenario to participants, or the perceived level of riskiness. 
            \\ *\emph{p}<0.05, **\emph{p}<0.01, ***\emph{p}<0.001}
            \label{ScenarioMANOVA}
\end{table}

\begin{table}[!htb] 
        \begin{tabular}{llllll}
            \toprule
            Factor              & DF & Pillai approx. & F & \emph{p}-value \\
            \midrule
            Scenario            & 3  & 0.076467 & 1.48413 & 0.18456 \\
            PRE Design          & 4  & 0.024725 & 0.35048 & 0.94493 \\
            Scenario:PRE Design & 12 & 0.302077 & 1.66049 & 0.03125* \\ 
            \bottomrule
        \end{tabular}
            \caption{Two-way MANOVA results for interaction model to explain any significantly different impacts of different tool and scenario combinations on perceived relatability or riskiness of scenarios. 
            \\ *\emph{p}<0.05, **\emph{p}<0.01, ***\emph{p}<0.001}
            \label{ScenarioToolTestxScenarioRanks}
\end{table}

\begin{table}[ht]
\centering
\begin{tabular}{lcccc}
\toprule
Fixed Effect & Estimate ($\beta$) & Std. Error & t value \\
\midrule
Intercept   & 5.477 & 0.218 & 25.184 \\
Scenario2   & 0.341 & 0.266 & 1.280 \\
Scenario3   & -0.310 & 0.326 & -0.950 \\
Scenario4   & -0.228 & 0.349 & -0.653 \\
\midrule
\multicolumn{4}{l}{\textit{Random Effects:} PID (Intercept) variance = 0.521, residual variance = 1.561} \\
\multicolumn{4}{l}{Number of observations = 132; number of participants = 44} \\
\multicolumn{4}{l}{F-test for Scenario: $F(3, \text{df}) = 1.790$, Sum Sq = 8.379} \\
\bottomrule
\end{tabular}
\caption{Linear mixed-effects model predicting $S\_ConcernRelatable$ from Scenario with random intercepts for participants (PID).}
\label{tab:lmer_results}
\end{table}

\begin{table}[ht]
\centering
\begin{tabular}{lccc}
\toprule
Fixed Effect & Estimate ($\beta$) & Std. Error & t value \\
\midrule
Intercept   & 5.886 & 0.172 & 34.149 \\
Scenario2   & -0.273 & 0.190 & -1.434 \\
Scenario3   & -0.172 & 0.236 & -0.728 \\
Scenario4   & -0.544 & 0.253 & -2.151 \\
\midrule
\multicolumn{4}{l}{\textit{Random Effects:} PID (Intercept) variance = 0.511, residual variance = 0.796} \\
\multicolumn{4}{l}{Number of observations = 132; number of participants = 44} \\
\multicolumn{4}{l}{F-test for Scenario: $F(3, \text{df}) = 1.694$, Sum Sq = 4.046} \\
\bottomrule
\end{tabular}
\caption{Linear mixed-effects model predicting $S\_Risky$ from Scenario with random intercepts for participants (PID).}
\label{tab:lmer_results_risky}
\end{table}

\subsection{Results of Interaction Effect Between Scenario and PRE Design on PRE Design Helpfulness and Ability to Address Privacy Concerns}
\begin{table}[!htb] 
        \begin{tabular}{llllll}
            \toprule
            Factor              & DF & Pillai approx. & F & \emph{p}-value \\
            \midrule
            Scenario            & 3  & 0.073098 & 1.4163 & 0.2093 \\
            PRE Design          & 4  & 0.079404 & 1.1576 & 0.3261 \\
            Scenario:PRE Design & 12 & 0.214087 & 1.1188 & 0.3243 \\ 
            \bottomrule
        \end{tabular}
            \caption{Two-way MANOVA results for interaction model to explain any significantly different impacts of different tool and scenario combinations on perceived helpfulness of ability for tool to address users concerns. 
            \\ *\emph{p}<0.05, **\emph{p}<0.01, ***\emph{p}<0.001}
            \label{ScenarioToolTestxDesignRanks}
\end{table}

\begin{table}[h!]
\centering

\begin{tabular}{lccc}
\toprule
Fixed Effect & Estimate & Std. Error & t value \\
\midrule
(Intercept)          & 5.3603   & 0.4761 & 11.259 \\
Scenario2            & 1.0114   & 0.6942 & 1.457  \\
Scenario3            & -1.6560  & 1.0577 & -1.566 \\
Scenario4            & -0.6082  & 0.8200 & -0.742 \\
Tool2                & -0.5333  & 0.6935 & -0.769 \\
Tool3                & 0.1481   & 0.6120 & 0.242  \\
Tool4                & 0.7128   & 0.7232 & 0.986  \\
Tool5                & 0.2568   & 0.6219 & 0.413  \\
Scenario2:Tool2      & 0.3038   & 0.9654 & 0.315  \\
Scenario3:Tool2      & 2.6752   & 1.3387 & 1.998  \\
Scenario4:Tool2      & 1.3506   & 1.1466 & 1.178  \\
Scenario2:Tool3      & -0.6153  & 0.9470 & -0.650 \\
Scenario3:Tool3      & 1.5896   & 1.2380 & 1.284  \\
Scenario4:Tool3      & 0.6357   & 1.3393 & 0.475  \\
Scenario2:Tool4      & -1.0761  & 1.0154 & -1.060 \\
Scenario3:Tool4      & -0.6033  & 1.3311 & -0.453 \\
Scenario4:Tool4      & 0.2046   & 1.1166 & 0.183  \\
Scenario2:Tool5      & -1.7757  & 0.9239 & -1.922 \\
Scenario3:Tool5      & 2.0853   & 1.2960 & 1.609  \\
Scenario4:Tool5      & -1.3334  & 1.3196 & -1.011 \\
\bottomrule
\end{tabular}
\caption{Linear mixed-effects model results for $S\_ConcernRelatable$. Random intercepts for participants included. \textit{Random effects:} Participant intercept variance = 0.3843, Residual variance = 1.5258.}
\label{tab:LMM_S_ConcernRelatable}
\end{table}

\begin{table}[h!]
\centering
\begin{tabular}{lccc}
\toprule
Fixed Effect & Estimate & Std. Error & t value \\
\midrule
(Intercept)          & 6.2619   & 0.3684 & 16.996 \\
Scenario2            & -0.7243  & 0.5277 & -1.373 \\
Scenario3            & -0.6114  & 0.7997 & -0.765 \\
Scenario4            & -0.8205  & 0.6212 & -1.321 \\
Tool2                & -0.6774  & 0.5262 & -1.287 \\
Tool3                & -0.4344  & 0.4651 & -0.934 \\
Tool4                & -0.4992  & 0.5483 & -0.910 \\
Tool5                & -0.3248  & 0.4708 & -0.690 \\
Scenario2:Tool2      & 0.8661   & 0.7413 & 1.168  \\
Scenario3:Tool2      & 0.8277   & 1.0266 & 0.806  \\
Scenario4:Tool2      & 0.8784   & 0.8746 & 1.004  \\
Scenario2:Tool3      & 0.5921   & 0.7331 & 0.808  \\
Scenario3:Tool3      & 0.1408   & 0.9378 & 0.150  \\
Scenario4:Tool3      & 1.3676   & 1.0319 & 1.325  \\
Scenario2:Tool4      & 0.2212   & 0.7799 & 0.284  \\
Scenario3:Tool4      & 1.1649   & 1.0048 & 1.159  \\
Scenario4:Tool4      & 0.4260   & 0.8504 & 0.501  \\
Scenario2:Tool5      & 0.5482   & 0.7103 & 0.772  \\
Scenario3:Tool5      & 0.1453   & 0.9885 & 0.147  \\
Scenario4:Tool5      & -1.6384  & 1.0004 & -1.638 \\
\bottomrule
\end{tabular}
\caption{Linear mixed-effects model results for $S\_Risky$. Random intercepts for participants included \textit{Random effects:} Participant intercept variance = 0.5741, Residual variance = 0.7765. }
\label{tab:LMM_S_Risky}
\end{table}

\begin{table}[!htb] 
        \begin{tabular}{llllll}
            \toprule
            Factor                  & Mean & SD & Skew \\
            \midrule
            Scenario Riskiness      & 5.62  & 1.23 & -0.98 \\
            Scenario Relatability   & 5.52  & 1.41 & -1.36 \\
            \textbf{Mean Inter-Item Correlation = \textit{0.237}} \\
            \midrule
            PRE Helpfulness         & 5.87  & 1.11 & -1.22 \\
            PRE Addressed Concern   & 5.51  & 1.17 & -0.95 \\
            \textbf{Mean Inter-Item Correlation = \textit{0.226}}  \\
            \bottomrule
        \end{tabular}
            \caption{Assessments of mean inter-item correlation of the scale items assessing the perceived risk level of the scenario, the relatability of the scenario (0.226, 0.237) were found to fit within the acceptable range (0.15-0.50) of inter-item correlation, showing that the scale items correlated well enough that they measure the same concept but were not so similar so as to be repetitive.}
            \label{InterItemCorr}
\end{table}

\clearpage

    \onecolumn
\small
\section{Qualitative Analyses}
\subsection{Codebook + examples}
\begin{longtable}{
  @{}
  >{\small\selectfont\raggedright\arraybackslash}p{0.2\textwidth}
  >{\small\selectfont\raggedright\arraybackslash}p{0.3\textwidth}
  >{\small\selectfont\raggedright\arraybackslash}p{0.02\textwidth}
  >{\small\selectfont\raggedright\arraybackslash}p{0.45\textwidth}
  @{}
}
\caption{Codebook - Single Codes} 
\label{codebook} \\
\toprule
&Description & N & Example Quotes    \\ 
\midrule
\endfirsthead

\toprule
                                    &Description            & N                    & Example Quote    \\ 
\midrule
\endhead

\midrule 
\multicolumn{4}{r}{{Continued on next page}} \\
\endfoot

\bottomrule
\endlastfoot

        \textbf{Awareness Reaction (AR)} & The character’s immediate emotional response that arises upon seeing the tool’s initial output or by the situation itself. & ~ & ~ \\ \hline
        AR: PER negative & The character expresses a negative emotional reaction towards the situation. & 17 & “She is scared and worried that someone she knows might figure out it is her” (P41). \newline
        
“He is worried of the risks that online communities can be pretty toxic and is worried that their posts will piss off more extreme people to the extend that they will troll, harass or even attempt to doxx Gray” (P26). \\ \hline
        AR: Coupled w/Tool Use: Positive & Tool output elicits a positive (e.g. reassuring, calming, empowering) emotional response from the character. & 6 & “Mel is likely feeling a significant sense of relief and a boost in confidence regarding their anonymity” (P14). \newline
        
“Emma feels relieved that this technology exists. She will now follow the advisement of the scanning tool” (P20). \\ \hline
        AR: Coupled w/Tool Use: Negative & Tool output elicits a negative (e.g. anxiety, fear, shame, frustration, distress) emotional response from the character. & 63 & “Ren feels uneasy about how her posts reveals, especially with the tool showing a 35\% risk of being identified” (P47).  \newline
        
“Ren's concerns about keeping her identity private appear to have been heightened by the privacy assessment results displayed in panel 4” (P23). \\ \hline
        \textbf{Risk Perception (RP)} & Whether the character’s risk perception is shaped primarily by the tool or by their own judgment. & ~ & ~ \\ \hline
        RP: Tool derived & The character draws their perception of risk directly from the tool’s output. They take the tool’s output “at its word”. & 78 & Seeing that some phrases increased his risk by over 30\% made him feel exposed, even before hitting “Post.” The tool made him question whether he could share his story safely at all” (P44). \newline
        
The tool pointed out to Ren that her post had a moderate risk because she said her family members, age, relationship status, and location when put together make it likely someone could identify her” (P12). \\ \hline
        RP: User Derived (grain of salt) & The character forms their own judgment about risk. They may override or question the tool’s assessment. & 31 & ""Though the tool only assigned a ""moderate"" privacy risk, this is probably too much of a risk- more than she should be willing to attempt” (P17). \newline
        
""Mel thinks that moderate seems like too much of a risk"" (P33). \\ \hline
        \textbf{Concern Scope (CS)} & The scope of the character’s privacy concern—what specifically they are worried might reveal their identity or put them at risk. & ~ & ~ \\ \hline
        CS: Specific post
(default) & The character is only concerned about risks related to the specific Reddit post. & 122 & “She thinks her ex can identify her. Using the information provided on her post” (P14). \newline

“Mel edits his post using the tool’s suggestions and feels safer sharing it” (P44). \\ \hline
        CS: Other Account Info (e.g. Reddit username) & The character is concerned about information associated with their Reddit account that could reveal their identity. This can include the history of posts, usernames, profile data, or other linked content that might connect them to their offline identity. & 7 & “Alex advises setting up a new Reddit account with no personal information, utilizing a VPN, and being extremely watchful of the tone and content of her posts” (P23). \newline
        
“Emma knows that some posters have been identified, even when using throw-away accounts and she's hesitant about being discovered” (P15). \\ \hline
        CS: Other platforms & The character is concerned about being identified or linked across platforms outside of Reddit.
This can be in addition to the already assumed concern over Reddit. & 3 & “Over the following weeks and months, Gray implements the recommended privacy measures, such as using more secure communication channels, adjusting their online behaviour, and exploring alternative platforms that prioritize user anonymity. The collaboration with Alex proves invaluable, as they work together to address each issue identified in the assessment and ensure Gray's online presence remains protected.” (P23) \newline 

“It at least would make him aware of his digital footprint and what he was posted on line” (P46). \\ \hline
        \textbf{Threat Model (TM)} & Whom the character does NOT want to see the post. & ~ & ~ \\ \hline
        TM: Narrative-defined (default) & The threat model aligns with what is described in the storyboard. & 122 & “She thinks her ex can identify her” (P14). \newline 
        
“Gray hesitate to share it at first worrying that even small clues might be enough for a motivated troll to uncover their identity” (P21). \\ \hline
        TM: Participant Defined & The threat model is different from or extends beyond what is described in the storyboard. & 11 & “this could be seen by anyone when people shares the post and could make her lose her job hence her concerns” (P45). \newline
        
“After editing the post, Ren decides to post it. after few weeks a workmate recognizes the post brings the conversation about it in workplace group chat” (P43). \\ \hline
        \textbf{Perceived Repercussions (PR)} & ~ & ~ & ~ \\ \hline
        PR: Narrative-defined
(default) & Perceived repercussions align with what is described in the storyboard.  & 123 & “He is worried of the risks that online communities can be pretty toxic and is worried that their posts will piss off more extreme people to the extend that they will troll, harass or even attempt to doxx Gray” (P26).  \newline 

“If the privacy in the post is exposed, the ex will to Emma will be able to view it and might result to a serous issue” (P42). \\ \hline
        PR: Participant-Defined & Perceived repercussions are different from or extend beyond what is described in the storyboard. & 10 & “Little did she know that her brand new co-worker came across the post because, coincidentally, they also happen to be a moderator of the sub reddit r/jobs. The co-worker would love nothing more than to have Rens position at the facility where they work and so she wastes no time filling the boss in on Rens post” (P17). \newline
        
“Gray would have so many unwanted contacts from people who disagree about his political views. The consequences can be really bad, some people are loosing their families forever. This tool really saves lives as there is still so much violence in this world, although it is in the human nature as well to share your life and thoughts with others, we all want to stay safe and make it to another day” (P3). \\ \hline
        \textbf{Perspective Shift (PS)} & The change in how the character thinks about privacy, risk, or their own behavior. & ~ & ~ \\ \hline
        PS: Awareness of information inferrability & Participant indicates heightened awareness of the potential for specific pieces of information disclosed in the post to lead to the character’s re-identification. & 66 & “She realizes that even a small details can significantly increase her exposure” (P47). \newline 
        
“the tool indicate that certain parts of the post make them identifiable” (P21). \\ \hline
        PS: Awareness of Risk & The participant expresses the character gaining a clearer understanding that online disclosure could result in harm.  & 39 & “Based on the panel, Mel thinks that her risk of being identified is a significant risk” (P5). \newline
        
“She thinks that there is a high risk of being identified. There are only 20 people in the entire world who match her characteristics. That is a very low number. She is terrified of being found out” (P30). \\ \hline
        PS: Increased Awareness of Audience & The participant shows an increased awareness of the character’s exposure with respect to who might see the post. & 8 & “She now sees that some of her references or stories may ring a bell to people at her workplace, heightening her concerns about people figuring out her identity, and finding her, and her blog when they do” (P11). \newline
        
“She didn't know her post had so many details that could identify her to some people. Seeing that only 20 people matched her info made her feel more targeted than she thought” (P40). \\ \hline
        \textbf{Self-censorship enacted?} & Actions that the character takes in response to the risks surfaced. & ~ & ~ \\ \hline
        Extreme Self-Censorship: Delete After (unedited post) & The participant describes how the character makes unedited post, but later deletes the content due to privacy or safety concerns (retrospective correction). & 2 & “Because her post was so specific and could be identified down to 20 people in the entire world, she decides to delete her post. That is the much safer option” (P30). \newline
        
“When Gray sees the post rating, even if the range is 20 people and he feels concerned about being tracked down by extremists online or in person, he makes the conscious decision to remove the post and not edit or change it in any way” (P37). \\ \hline
        Extreme Self-Censorship: Delete After (edited post) & The participant describes how the character makes edited post, but later deletes the content due to privacy or safety concerns (retrospective correction). & 4 & “She removes most identifying information and comes up with a more anonymous approach to posting online. [...]She likely deletes the post” (P37). \newline
        
“Emma finnaly posts on reddit after all the edits. Weeks later, her ex finds out about the posts and sends her threatening messages. Panicked her takes down the posts and stops using the tool feeling let down with the tool” (P43). \\ \hline
        Extreme Self-Censorship: Doesn't Post & The participant describes the character choosing not to post at all. & 18 & “Gray decide not to go ahead with the post” (P25). \newline
“Emma ultimately decides against making a post on the subreddit for some assistance” (P27). \\ \hline
        Extreme Self-Censorship: Leaves platform & The participant describes how the character decides to leave Reddit. & 4 & “I think the best thing that Mel should do is opt for an offline discussion with trusted individuals instead of posting anything online that could be a risk” (P5). \newline
        
“I also think Emma would have chosen another platform to post her posts as posting on reddit may pose her privacy to so much risks” (P42). \\ \hline
        Moderate self-censorship: Edited Flagged Risks & The participant describes how the character specifically edits parts of the post that were highlighted as risky by the tool (e.g., changing or removing named entities or identifiable details). & 80 & “After rereading the flagged section, she rewrite her post in a way that generalizes key details while keeping her message” (P4). \newline
        
“Gray was able to redo her post and take out any information that pointed at her” (P7). \\ \hline
        No self-censorship: Posts Unedited & The participant describes how the character keeps the post as-is, without making any edits or removing content. & 9 & “Given the answer back that there was only about 20\% of people in the population who fit that criteria he was more at ease with leaving his post with out any changes” (P17). \newline
        
“He realizes the effort required to readjust his post would take too much time. He would do very minor corrections but 90\% of the post is the same as before. He would proceed with sending it” (P46). \\ \hline
        Action: Post edited, Other & The character edits the post in a way not directly related to self disclosure, such as changing tone, framing, or language to avoid negative response. & 1 & “Gray ultimately decides to make the post on the political topic, but decides to go about it in a way that they want to be as true to their beliefs as possible while remaining respectful of the views of others on the platform” (P27). \\ \hline
        Action: Not described/unclear & The participant does not clearly describe what action the character ultimately takes. & 18 & ~ \\ \hline
        \textbf{Outcome: Were they re-identified or not?} & Whether the character was ultimately re-identified due to the self-disclosures included in their post. & ~ & ~ \\ \hline
        Not re-identified (after posting) & Character makes post (either edited or unedited) and IS NOT re-identified. & 62 & “Gray was able to redo her post and take out any information that pointed at her” (P7). \newline
        
“she can reap the rewards and not be exposed to harm” (P24). \\ \hline
        Re-identified (after posting) & Character makes post (either edited or unedited) and IS re-identified. & 11 & “She thought she deleted enough information but then her worst nightmare came true. One of her coworkers figured out it was her” (P41). \newline
        
“Weeks later, A cowoker stumbles upon the post and ashares it with the HR. Suspecting Mel's involvement, mel is called into am meeting for questioning” (P43). \\ \hline
        Not re-identified (didn’t post) & Character does not make post and is NOT re-identified. & 18 & “she still feels uncertain and decided not to publish the post. instead she decided to bookmark a few resources that others had shared in similar thread and start drafting a more private message to a moderator asking if there is a safer way to ask for help. in the following week she learns more about privacy” (P21). \newline
        
“in this situation i think that gray ultimately decides not to post his thoughts on the matter in the forum after the tool showed him a higher chance of being identified, i think he just doesn't trust the tool but still wants to express his views so i think he will get offline and call up a friend and debate the topic” (P19). \\ \hline
        Unclear/Not described & It is not clearly described whether or not the character’s actions lead to re-identification. & 29 & ~ \\ \hline
        Anything could happen & The participant describes both outcomes of the character being re-identified and not being re-identified. & 10 & “The are only two scenarios in this case. First is she is likely to get caught by the boyfriend. Secondly she is not likely to get caught. We can not be sure of the outcome. However if she is caught by the boyfriend, things will not be in her favor and if she is not caught, things will be in his favor. Therefore i believe that this tool however much we do not know the outcome, it serves it's purpose in identifying whether or not she will be caught by the boyfriend. If it happens that she is not caught by the boyfriend, emma will be very happy and content but if she is caught, everything will work against her” (P9). \newline
        
“For the empty yellow panel, I think Emma will probably be confused and unsure how to change her post to be safer while still getting the advice she needs. She might try rewriting it a couple of times while feeling stuck and frustrated about what details to remove. She might even consider not posting it at all or decide to post a less detailed version and hope for the best. This experience could make her more cautious about sharing personal information online in the future. Its always important to weigh the potential benefits against the risks to personal privacy” (P5). \\ \hline
        \textbf{Outcome: Did the tool help them reap the benefits of self-disclosure online?}  & Whether the character was ultimately able to meet their original goals for self-disclosure (e.g. receiving advice, expression opinions), and whether the tool played a role in enabling that outcome. & ~ & ~ \\ \hline
        No, it didn’t & ~ & 29 & ~ \\ \hline
        No, it didn’t: Needs met, but not by tool (e.g. by external resource) & The character achieves their goals, but only through support outside of the tool. (e.g. receiving advice or support from a friend, therapist, or another platform) & 10 & “Over the following weeks and months, Gray implements the recommended privacy measures, such as using more secure communication channels, adjusting their online behaviour, and exploring alternative platforms that prioritize user anonymity. The collaboration with Alex proves invaluable, as they work together to address each issue identified in the assessment and ensure Gray's online presence remains protected” (P23). \newline
        
“'m think Ron ends up getting a friend or a mentor to talk to who might help her with whatever she's looking for in Reddit” (P34). \\ \hline
        No, it didn’t: Needs not met (posted + didn’t get desired reaction) & The character posts, but does not receive the support, response, or outcome they hoped for. & 6 & “Emma, desperate to share her story, decided to take the risk - against the advice of the tool - to make a post on Reddit. She just had to share her own side of the story and hoped the online world world would believe her. It so turned out that she was mocked and claimed to not be submissive or understanding enough which warranted the treatment she received from her ex. Just a minute percentage of people show any form of compassion or believed her. Besides, her Ex decided to come out and debunk all her narratives, even worse, calling a liar and accusing her of blackmail” (P36). \newline
        
“Weeks later, A cowoker stumbles upon the post and ashares it with the HR. Suspecting Mel's involvement, mel is called into am meeting for questioning, Mel denies the post, However they are put into probation, this hightened their distrust for the tool prompting them to seek futher for othe safer alternatives that could help without exposing their identities” (P43). \\ \hline
        No, it didn’t: Needs not met (they didn’t make the post) & The character decides not to post, and thus cannot achieve their goals of self-disclosure. & 13 & “Emma looked at the screen for few minutes, trying to weight her options. She really wants to reach out for help but doesn't wanna open because of embarrassment. After reading the privacy report a few times, She tries editing her post, replacing it with a specific city, name and date. she still feels uncertain and decided not to publish the post. instead she decided to bookmark a few resources that others had shared in similar thread and start drafting a more private message to a moderator asking if there is a safer way to ask for help. in the following week she learns more about privacy” (P21). \newline
        
“He decides that the risk isn't worth the post because it really isn't going to make a difference anyway. ‘Nobody really listens anymore’ he says to himself as he closes his laptop computer and drinks his unicorn slush from Sonic with quiet contemplation” (P29). \\ \hline
        Yes it did & ~ & 58 & ~ \\ \hline
        Yes it did: met (posted + got desired online reaction) & The character posts and receives the kind of reaction they were seeking (e.g. feeling validated, getting helpful responses, helping others by sharing). & 57 & “Later, she gets supportive responses on reddit from people on similar situations” (P4). \newline
        
“It turns out there are a lot of people who work in a similar field that have had the same issues with their supervisor - they were able to tell her about their experiences and outcomes which gave her steps to try in her own job” (P7). \\ \hline
        Yes it did: Needs met, but only after receiving outside support to understand tool outputs & The character benefits from the post, but only after getting help interpreting the tool (e.g., from a friend or external resource). & 1 & “Ren chooses to get more help after realizing how serious the situation is. In an attempt to obtain new insight and direction on handling this delicate circumstance, she contacts a mentor or close friend[...]She begins to create a strategy with the assistance of her confidante, one that strikes a compromise between her need for seclusion and the actions required to protect her internet reputation” (P23). \\ \hline
        Unclear/Not described & The participant does not clearly explain what ultimately happens or whether the character’s needs are met. & 46 & ~ \\ \hline
        Anything could happen & The participant describes multiple outcomes without stating whether the character's needs were ultimately met. & 11 & “In the empty yellow panel there two possibilites that may happen to Gray. Gray can decide to use the tool to avoid all the harassment from social media or he might decide not to use the tool and he will be harassed in the social media by the community which he does not like. I feel for Gray because social media harassment is so bad and can lead to depression if one is not kin about it. This is too bad for gray if this happens. Social media harassment can impact one life so badly. I really feel for gray” (P2). \newline
        
“In the empty yellow panel, there are probabilities of only two outcomes. First of all, there is a probability that the post she will post online will be know by her colleague if she does not follow what the tool advises her. That is; desist from making her location be known, her family members, her age and race. If she follows this, there is a very low chance that the post will be known. However if she does not follow the advise given by the tool, there is a very high probability that her post will be known and she faces the risk of loosing her job (P9). \\ \hline
        \textbf{Outcome: How do they use the tool (if at all?)} & Whether and how the character used the tool. & ~ & ~ \\ \hline
        Disregards tool & The character completely ignores or chooses not to engage with the tool in any capacity.
They do not use it to evaluate, reflect, or edit their post. & 5 & The fact that the information Mel provides matches the identity of only 20 people worldwide makes it a huge risk of posting the issue online especially as it involves exposing the company and bring its actions to the public eye. (RQ1) \newline 

Mel understands that the obstacles faced is enormous because she can be easily identified and going out against the company signals the beginning of the end for her job in the organization. Thus, the odds are stacked against her if she does decides to go ahead with her plan. (RQ3) 

Mel, against all odds, decided that speaking out against the unethical behaviors of the company is the right thing to do even if it costs her her job. It matters to her that organizations to have all their way around the business world even if they engaged in unethical activities that the law stand against. This prompted her decision to stand for the right thing. And thus, spoke out on the negative actions of the company in as much as the odds are stacked against her. She did not bother or flinch she was going out against the company rather she was bold as ever for standing for the right thing to do. (RQ2)
-(P36, Pos. 3-5) \\ \hline
        For modifying post & The character actively uses the tool’s feedback to guide modifications to their post. & 87 & ~ \\ \hline
        For risk awareness at minimum & The character acknowledges the risk surfaced by the tool and makes a decision influenced by that awareness, but does not necessarily edit or modify the post.
This includes cases where the character:
Does not post at all due to the highlighted risks; Engages reflectively with the tool’s feedback but chooses not to change the post & 34 & “She is afraid and at the same time she thinks it's a good idea to get advice but she is not free to post. His ex might embarrass her too like others who have been going through the same experience as her, her being concern is being helped out with her problems but being afraid of the consequences she is left with many questions weather to post or not, maybe her problems will be more than others with similar problems to her. She downloads an app that will help her to know who can view her posts just to be safe and aware of what she may face” \\ \hline
        Unclear/Not described & The participant does not clearly describe whether or how the character engages with the tool. & 4 & ~ \\ \hline
        Anything could happen & The participant describes multiple ways the character uses the tool. & 8 & ~ \\ \hline
\end{longtable}
        
        \small
        
\clearpage
\onecolumn
\begin{longtable}{
  @{}
  >{\small\selectfont\raggedright\arraybackslash}p{0.2\textwidth}
  >{\small\selectfont\raggedright\arraybackslash}p{0.3\textwidth}
  >{\small\selectfont\raggedright\arraybackslash}p{0.02\textwidth}
  >{\small\selectfont\raggedright\arraybackslash}p{0.45\textwidth}
  @{}
}
\caption*{Codebook - Multiple Codes} \\
\toprule
&Description                    & N              & Example Quotes    \\ 
\midrule
\endfirsthead

\toprule
                                    &Description       & N                 & Example Quote    \\ 
\midrule
\endhead

\midrule 
\multicolumn{4}{r}{{Continued on next page}} \\
\endfoot

\bottomrule
\endlastfoot
        \textbf{Motivation} & The underlying reason for why the character is using the tool or taking a particular action. & ~ & ~ \\ \hline
        Motivation: Vulnerability (first hand harms) & The character is motivated to avoid harm to themselves. & 88 & “Ren decides to edit her post so that she is hopefully unidentifiable” (P41). \newline
        
“Emma thinks she has to alter her post to make herself less identifiable. She does not want her ex to see and harass her more” (P33). \\ \hline
        Motivation: Risk Reduction, Second hand harms & The character is motivated to avoid harm to others. & 4 & “I think Grays family is an obstacle because they might be dragged into this yet they have nothing to do about it” (P34).  \newline
        
“Emma feels her family members, age, location and relationship status is at risk of being exposed and the possibility of her ex who has been harassing her seeing it and the situation escalating into a worst situation” (P18). \\ \hline
        Motivation: Privacy Fatigue & The character displays emotional or cognitive fatigue in response to privacy demands/decisions. & 8 & “Frustrated and exhausted, Mel closes their laptop without posting anything, feeling silenced and alone” (P35). \newline
        
“Gray drafts and deletes and drafts and deletes multiple versions of the post. Each time, the posts became less and less clear and more and more vague. Ultimately, they just deleted the post; the post reduced to a fractured collection of messy, unrealized ideas that would never be shared” (P11). \\ \hline
        Motivation: Tool Goalposting & The character is motivated to meet the tool’s standards (e.g. to get a good risk score). & 16 & “She would then try to work on fixing those lines to make sure they don't appear as red. Once she has fixed them, her anxiety would go down” (P39). \newline
        
“After revising and seeing how a lower risk level from the tool, she posts” (P10). \\ \hline
        Motivation: Other & The character’s motivation is clearly expressed but does not fit into any of the above categories. & 2 & “Mel, against all odds, decided that speaking out against the unethical behaviors of the company is the right thing to do even if it costs her her job” (P36). \newline
        
“Although she realizes that she could possibly be identified having this information out there, it is critical for readers to know specifically what she does at her job and the location of it for them to understand the issue she is currently dealing with. She decides to leave it anyway” (P17). \\ \hline
        \textbf{User Agency (UA)} & The character’s feeling of agency in navigating decisions around posting. & ~ & ~ \\ \hline
        UA: (Friction) Balancing Post & The participant describes the character’s effort to figure out how to post in a way that balances risk and meaningful/authentic expression. & 58 & “She has some difficulty rewriting the post without the information in it as it doesn't sound that good anymore” (P16). \newline
        
“She needs to remove some information but some of that information is key to her story and post. She is having a hard time” (P41). \\ \hline
        UA: (Friction) Whether to post & The participant describes the character’s internal conflict around whether to post at all, based on weighing benefits of self-disclosure against re-identification risks. & 21 & “Although Ren is facing difficult challenges at her workplace in connection to her boss, she decides against posting online. She had weighed her options and considered if it is actually worth risking her job considering the fact that someone had earlier done what she intend doing and lost her job” (P36). \newline
        
“Gray reads the privacy tool after assessing his draft reddit post. He sighs wondering the potential risks to himself and his family are worth the political post” (P29). \\ \hline
        UA: Disempowerment/Overwhelm & The participant describes the character feeling stuck, overwhelmed, or unsure how to proceed due to the weight of the decision or lack of guidance. & 29 & “She might be second-guessing her decision to share certain aspects of her experience and feeling vulnerable knowing how easily her ex could potentially piece together her identity. The urgency to understand and mitigate this risk would likely be paramount in her mind” (P14). \newline
        
“Mel feels overwhelmed by the number of flagged items and is unsure which details are most crucial to change to effectively reduce their risk” (P35). \\ \hline
        UA: Empowerment & The participant describes the character feeling more confident, secure, or capable of posting as a result of engaging with the tool. & 34 & “While the tool doesn't give precise guidance, the act of editing helps Mel feel more secure” (P47). \newline
        
“Now she can freely post stories and ask for advice without feeling like she will be exposed” (P12). \\ \hline
        \textbf{Parallel Action(s)} & Additional actions the character takes outside of editing the post itself, in parallel to—or as a result of—the risks highlighted by the tool. & ~ & ~ \\ \hline
        Parallel Action: Off Reddit & The participant describes how the character takes an action outside of Reddit in response to the risk raised by the tool. & 20 & “Emma chooses to ask her trusty friend Alex for advice after she has been staring at the computer for what seems like an eternity. Alex assists in deciphering the technical output and outlining its main consequences as they carefully go over it together. Emma can take certain actions to lessen the risks, even when the assessment identifies certain possible weaknesses. Alex advises setting up a new Reddit account with no personal information, utilizing a VPN, and being extremely watchful of the tone and content of her posts.  \newline
        
Emma, feeling a little more in control, starts to make plans. She will take great care when crafting her Reddit post, steering clear of any personal references or identifying information. She will also closely monitor the activity, prepared to respond quickly if she notices any questionable conduct or tries to identify the sender.” \\ \hline
        Parallel Action: On Reddit & The character takes actions within the Reddit platform other than editing the single post. & 10 & “Mel could also find a smaller sub, r/jobs is a massive and general sub and maybe a smaller sub like ""r/ethics"" might be better” (P15). \newline
        
“She will also make sure that her Reddit username is either anonymous or changed to something that her ex wouldn't recognize as well” (P17). \\ \hline
        \textbf{Residual Feelings (RF)} & Feelings that remain after the character has taken action (or chosen inaction). & ~ & ~ \\ \hline
        RF: Concern (Negative) & Even after making changes or using the tool, the character still has concerns about potential re-identification or consequences. & 22 & “The moderate rating leaves them feeling somewhat uncertain. It's not a definitive high risk, but it's enough to warrant caution” (P1). \newline
        
“i think she will be uncertain if the score is sufficient enough to keep her anonymous” (P19). \\ \hline
        RF: False Sense of Security/Overreliance (Negative) & The participant implies or explicitly states that the character places too much trust in the tool, potentially underestimating real risks. & 5 & “Panicked her takes down the posts and stops using the tool feeling let down with the tool” (P43). \newline
        
“Emma thinks nothing of it because the tool scanned her post as having a very low privacy risk. But she fails to change her Reddit username- something that her ex has remembered- and he has now found her post and is aware of her plans for later in the week” (P17). \\ \hline
        RF: Tool referral (Positive) & The character expresses the intention to recommend the tool to others. & 8 & “Inspired by this, Gray begins to advocates for clearer, more user friendly privacy tools and becomes part of feedback groups for the tool developers” (P4). \newline
        
“Later, Ren shares the tool with a friend going through similar situation, happy she found a way to speak up within risking her job or reputation” (P10). \\ \hline
        \textbf{User Experience Related Codes} & These codes describe participants’ reflections on how the characters experienced the tool.  & ~ & ~ \\ \hline
        \textbf{Interpretability} & How well the participant/character understands or struggles to understand the internal logic/mechanics of the tool itself.
E.g., how risk scores are generated or how the model determines what is risky. & ~ & ~ \\ \hline
        Interpretability (Con) & The participant/character is unclear on how the tool works or how it produces its outputs. \newline 
        E.g. Seeing the tool as a ""black box"; expressing confusion about how scores or risk assessments are calculated; “Where is this coming from?”; Wanting more understanding around how the tool makes its decisions and not just what it flags & 17 & “She rereads the flagged categories, trying to decipher how seemingly innocuous details in her story could be pieced together by her ex” (P1). \newline
        
“The tool uses a ""k-anonymity score"" that supposedly calculates how many other people in the world share the traits they described in their post” (P13). \\ \hline
        Interpretability (Pro) & The participant/character demonstrates an understanding of how the tool produces its outputs. \newline
        E.g. Correctly describing the logic behind how the tool evaluates risk; Demonstrating an accurate mental model of how the tool works; Acknowledges clarity of the tool’s internal process
          & 3 & “Quantifiable data sets are always good empirical evidence that can belay emotional concerns” (P29). \newline

“Emma would first go through all the information to understand how the tool works” (P39). \\ \hline
        \textbf{Ability} & How easily the character is able to enact action that addresses privacy risks and enables disclosure benefits. & ~ & ~ \\ \hline
        Ability (Con): Needs Guidance & The participant/character implicitly or explicitly expresses a desire for help or clearer direction on how to revise or de-risk their post. \newline
        E.g. Wanting specific suggestions/automation; Struggling with vague feedback; Feeling burdened by the responsibility to interpret and act alone & 55 &  “Ren spends a long time re writing her post, trying to guess what tools sees as risky, frustrated and unsure if her post is safe, she eventually decides not to share it at all “ (P21). \newline

“The tool identifies the risky elements but doesn't provide concrete advice on how to rephrase her thoughts” (P1). \\ \hline
        Ability (Pro): Balanced posting & The participant/character feels that the tool supports their ability to post safely and meaningfully.
Achieving a balance between self-expression and risk reduction & 4 & “She goes back to her message and removes the specifics like her age and gender, while still maintaining the core message of her post so she is able to get the best quality of responses possible” (P13). \newline

“She types the passage again adding minimal amount of information to supply necessary context” (P29). \\ \hline
        \textbf{Explainability} & How well the participant can understand what the tool is showing or telling them. & ~ & ~ \\ \hline
        Unclear Meaning (Explainability Con) & The participant/character finds the tool’s output or underlying concepts difficult to understand. & 39 & “Emma is confused about whether the assessment means she is more or less likely to be identified at first based off of the wording” (P33). \newline

        “The dense, data-heavy information presented in the assessment likely requires a high level of expertise that Gray may not possess” (P23).
        \\ \hline
        Not Specific Enough/Too General (Explainability Con) & The participant/character finds the tool’s feedback too vague or generalized to be actionable. & 31 & “Ren faces a difficult decision in understanding what the tool has identified and trying to get her point across in the post” (P17). \newline

        “She finds it difficult to understand how the percentage of risk score is calculated and which specific combinations of details are the most dangerous” (P47).
        \\ \hline
        Misinterpretation (Explainability Con) & The participant/character misinterprets the tool’s output or intended function/purpose. & 14 & “The panel states, ‘Combined, this information matches the identity of only 1 in every 20 people worldwide. This greatly reduces your re-identification risk’  Seeing that their combination of age, name, exact time frame, and location details only uniquely identifies them within a relatively small fraction of the global population would likely alleviate their initial worries about being identified by their workplace” (P14). \newline

        “When Ren uses the downloaded post, some of the issues that are affecting him in the job place with his boss will not be disclosed” (P38).
        \\ \hline
        \textbf{Transparency} & The participant/character questions whether the tool can be trusted, or raises concerns about hidden limitations or blind spots. This may include: Expressing general mistrust or skepticism; Wondering if the model is thorough, fair, or complete & 14 & “i think she will be uncertain if the score is sufficient enough to keep her anonymous. I think she will go back and fourth in her mind and ultimately decide to post in the forum after using the tool” (P19). \newline
        
“She also worries the tool might not catch everything, leaving her vulnerable” (P43). \\ \hline
    \end{longtable}

    \clearpage

\subsection{Co-Occurrence of Codes}
    \subsection{RQ1 Code Frequencies, \& Co-Occurrences}

\begin{table}[!ht]
\caption{RQ1: Co-Occurrence of Emotional Reactions to Population Risk Estimates with Perspective Shift}
\label{Table: ERxPS}
    {%
    \begin{tabular}{|l|l|l|l|}
    \hline
         & Awareness of Information Inferrability & Awareness of Risk & Increased Awareness of Audience \\ \hline
        Awareness Reaction Coupled w/Tool Use: POSITIVE  & 3 & 1 & 0 \\ \hline
        Awareness Reaction Coupled w/Tool Use: NEGATIVE  & 38 & 19 & 8 \\ \hline
    \end{tabular}
    }
\end{table}

\begin{table}[!ht]
\caption{RQ1: Occurrence of Perspective Shift Across Reflections}
\label{Table:PS}
    \centering
    \begin{tabular}{|l|l|l|}
    \hline
        Perspective Shift Sub-themes  & Occurrences Across Reflections & Percentage (\%) \\ \hline
        Awareness of Information Inferrability & 65 & 49.24 \\ \hline
        Awareness of Risk  & 39 & 29.55 \\ \hline
        Increased Awareness of Audience & 8 & 6.06 \\ \hline
        Total Number of Reflections Noting Perspective Shift & 98/132 & 74.24 \\ \hline
    \end{tabular}
\end{table}

\begin{table}[!ht]
\caption{RQ1: Risk Perception Sub-theme Frequencies}
\label{Table:RP}
    \centering
    \begin{tabular}{|l|l|l|}
    \hline
        Code System  & Occurrences Across Reflections & Percentage (\%) \\ \hline
        Perception of Risk is Tool derived  & 100 & 75.76 \\ \hline
        Perception of Risk is User Derived & 32 & 24.24 \\ \hline
    \end{tabular}
\end{table}

\begin{table}[!ht]
\caption{RQ1: Threat Model \& Perceived Repercussions}
\label{Table:TM_PR}
    \centering
    \begin{tabular}{|l|l|l|}
    \hline
        Code System  & Occurrences Across Reflections (N=132) & Percentage of occurrence Across Reflections (\%) \\ \hline
        Perceived Repercussions: Narrative-defined & 122 & 92.42 \\ \hline
        Perceived Repercussions: Participant-Defined & 10 & 7.58 \\ \hline
        Threat Model: Narrative-defined & 121 & 91.67 \\ \hline
        Threat Model: Participant Defined & 11 & 8.33 \\ \hline
    \end{tabular}
\end{table}

\clearpage

\begin{table}[!ht]
\caption{RQ1: Emotional Reaction to Population Risk Estimate Co-Occurrences with User Agency \& Misinterpretations}
\label{Table:ERxUA}
    \centering
    \begin{tabular}{|l|l|l|}
    \hline
        Code System & Awareness Reaction Coupled w/Tool Use: POSITIVE & Awareness Reaction Coupled w/Tool Use: NEGATIVE \\ \hline
        Difficulty Balancing Post & 0 & 33 \\ \hline
        Debating Whether to Post & 0 & 11 \\ \hline
        Dis-empowerment & 0 & 21 \\ \hline
        Empowerment & 3 & 20 \\ \hline
        Explainability Issue: Misinterpretation of Output & 2 & 2 \\ \hline
    \end{tabular}
\end{table}

\begin{table}[!ht]
\caption{RQ1: Risk Perception Co-Occurrences}
\label{Table:RPxPSxOIxTrans}
    \centering
    \begin{tabular}{|l|l|l|l|}
    \hline
        Code System & Perception of Risk is Tool derived (N=100) & Perception of Risk is User Derived (N=32) \\ \hline
        Awareness of Information Inferrability & 49 & 16 \\ \hline
        Awareness of Risk  & 28 & 11 \\ \hline
        Increased Awareness of Audience & 8 & 0 \\ \hline
        Transparency Issue & 6 & 10 \\ \hline
        Interpretability Issue & 11 & 6 \\ \hline
        Overall Impression: Positive & 48 & 10 \\ \hline
        Overall Impression: Mixed & 30 & 8 \\ \hline
        Overall Impression: Negative & 16 & 11 \\ \hline
        Overall Impression: Unclear & 6 & 3 \\ \hline
    \end{tabular}
\end{table}

    \subsection{RQ2 Code Frequencies, \& Co-Occurrences}

\begin{table}[!ht]
\caption{RQ2: Co-Occurrence of User Agency Codes with Motivation}  
\label{Table: UAxMotivation}
    \centering
    \begin{tabular}{|l|l|l|l|l|}
    \hline
         & Motivation: Vulnerability & Motivation: Tool Goalposting & Motivation: Privacy Fatigue  & Motivation: Other  \\ \hline
        Difficulty Balancing Post & 36 & 9 & 2 & 1 \\ \hline
        Debating Whether to Post & 14 & 0 & 5 & 1 \\ \hline
        Dis-empowerment & 16 & 3 & 7 & 0 \\ \hline
        Empowerment & 34 & 9 & 0 & 0 \\ \hline
    \end{tabular}
\end{table}

\clearpage

\begin{table}[!ht]
\caption{RQ2: Co-Occurrence of Population Risk Estimate Issues with Motivation}
\label{Table: UXIssuesxMotivation}
    \centering
    {%
    \begin{tabular}{|l|l|l|l|l|}
    \hline
         & Motivation: Vulnerability & Motivation: Tool Goal-posting & Motivation: Privacy Fatigue  & Motivation: Other  \\ \hline
        Interpretability & 14 & 0 & 2 & 0 \\ \hline
        Ability: Needs More Guidance & 35 & 10 & 7 & 0 \\ \hline
        Ability: Provides Balanced posting & 3 & 1 & 0 & 0 \\ \hline
        Explainability: Not Specific Enough & 19 & 3 & 3 & 1 \\ \hline
        Explainability: Misinterpretation of Output & 9 & 1 & 1 & 0 \\ \hline
        Explainability: Unclear Meaning & 25 & 5 & 5 & 0 \\ \hline
        Transparency & 12 & 2 & 0 & 0 \\ \hline
    \end{tabular}
    }
\end{table}

\begin{table}[!ht]
\caption{RQ2: Co-Occurrence of Enacted Self-Censorship with Motivation}
\label{Table:SCxMotivation}
    \centering
    {%
    \begin{tabular}{|l|l|l|l|l|}
    \hline
        Code System & Motivation: Vulnerability & Motivation: Tool Goalposting & Motivation: Privacy Fatigue  & Motivation: Other  \\ \hline
        Extreme Self-Censorship: Delete After & 2 & 1 & 0 & 0 \\ \hline
        Extreme Self-Censorship: Doesn't Post & 13 & 0 & 5 & 0 \\ \hline
        Extreme Self-Censorship: Leaves platform & 4 & 0 & 0 & 0 \\ \hline
        Moderate Self-Censorship: Edits Post & 61 & 15 & 0 & 0 \\ \hline
        No Self-Censorship: Posts Unedited  & 5 & 0 & 1 & 2 \\ \hline
        Other: Unclear & 9 & 0 & 2 & 0 \\ \hline
    \end{tabular}
    }
\end{table}

\begin{table}[!ht]
\caption{RQ2: Co-Occurrence of Risk Perception Codes with Motivation}  
\label{Table: MotivxRP}
    \centering
    {%
    \begin{tabular}{|l|l|l|l|l|}
    \hline
         & Motivation: Vulnerability & Motivation: Tool Goal-Posting & Motivation: Privacy Fatigue  & Motivation: Other  \\ \hline
        Perception of Risk is Tool Derived & 65 & 13 & 6 & 0 \\ \hline
        Perception of Risk is Participant Derived & 26 & 3 & 2 & 2 \\ \hline

    \end{tabular}
    }
\end{table}
    \subsection{RQ3 Code Frequencies, \& Co-Occurrences}

\begin{table}[!ht]
\caption{RQ3: User Agency and Re-identification Outcome Co-Occurrences}
\label{Table: UAxRe-ID Outcome}
    \centering
    \begin{tabular}{|l|l|l|l|l|l|}
    \hline
         & Not re-identified & Didn’t post & Re-identified & Outcome Unclear & Not Described \\ \hline
        User Agency: & 0 & 0 & 0 & 0 & 0 \\ \hline
         Difficulty Balancing Post & 21 & 6 & 6 & 6 & 17 \\ \hline
        Debating Whether to post & 5 & 10 & 1 & 2 & 3 \\ \hline
        Dis-empowerment & 8 & 11 & 0 & 1 & 8 \\ \hline
        Empowerment & 40 & 0 & 0 & 1 & 5 \\ \hline
    \end{tabular}
\end{table}

\clearpage

\begin{table}[!ht]
\caption{RQ3: Residual Feelings and Re-identification Outcome Co-Occurrences}
\label{Table: RFxRe-ID Outcome}
    \centering
    \begin{tabular}{|l|l|l|l|l|l|}
    \hline
         & Not re-identified & Didn’t post & Re-identified & Outcome Unclear & Not Described \\ \hline
        Residual Concern (Neg) & 8 & 4 & 5 & 1 & 4 \\ \hline
        False Sense of Security (Neg)   & 1 & 1 & 3 & 0 & 0 \\ \hline
        Tool referral (Pos)  & 8 & 0 & 0 & 0 & 0 \\ \hline
    \end{tabular}
\end{table}

\begin{table}[!ht]
\caption{RQ3: General Reaction and Re-identification Outcome Co-Occurrences}
\label{Table: UsedToolxRe-ID Outcome}
    \centering
    \begin{tabular}{|l|l|l|l|l|l|}
    \hline
        Code System & Not re-identified & Didn’t post & Re-identified & Outcome Unclear & Not Described \\ \hline
        Used for modifying post   55 & 3 & 6 & 1 & 20 & ~ \\ \hline
        Used for Risk Awareness (at minimum) & 7 & 15 & 2 & 3 & 6 \\ \hline
        Disregards tool & 0 & 0 & 3 & 0 & 2 \\ \hline
        Other: Not described & 1 & 0 & 0 & 0 & 1 \\ \hline
        Other: Unclear & 0 & 0 & 0 & 7 & 1 \\ \hline
    \end{tabular}
\end{table}

\begin{table}[!ht]
\caption{RQ3: Self-Censorship Reaction and Re-identification Outcome Co-Occurrences}
\label{Table: SelfCensorshipxRe-ID Outcome}
    \centering
    \begin{tabular}{|l|l|l|l|l|l|}
    \hline
        & Not re-identified & Didn’t post & Re-identified & Outcome Unclear & Not Described \\ \hline
        Extreme Self-Censorship: Delete After & 3 & 1 & 2 & 0 & 0 \\ \hline
        Extreme Self-Censorship: Doesn't Post  & 0 & 15 & 0 & 1 & 1 \\ \hline
        Extreme Self-Censorship: Leaves platform & 0 & 3 & 0 & 0 & 1 \\ \hline
        Moderate Self-Censorship: Edits Post & 53 & 0 & 5 & 1 & 19 \\ \hline
        No Self-Censorship: Posts Unedited  & 2 & 0 & 4 & 1 & 2 \\ \hline
        Other: Unclear & 5 & 0 & 1 & 6 & 6 \\ \hline
    \end{tabular}
\end{table}

\begin{table}[!ht]
\caption{RQ3: Other Reactions and Re-identification Outcome Co-Occurrences}
\label{Table: PAxRe-ID Outcome}
    \centering
    {%
    \begin{tabular}{|l|l|l|l|l|l|}
    \hline
        & Not re-identified & Didn’t post & Re-identified & Outcome Unclear & Not Described \\ \hline
        Off Reddit: Research advice & 1 & 1 & 0 & 0 & 0 \\ \hline
        Off Reddit: Seek Peer feedback & 0 & 5 & 0 & 1 & 3 \\ \hline
        Off Reddit: Enacting S\&P Measures & 2 & 0 & 0 & 0 & 1 \\ \hline
        Off Reddit: Seek Help to Interpret Population Risk Estimate & 0 & 2 & 2 & 0 & 2 \\ \hline
        On Reddit (e.g. change username, burner account, etc.) & 4 & 1 & 1 & 0 & 4 \\ \hline
    \end{tabular}
    }
\end{table}

\clearpage

\begin{table}[!ht]
\caption{RQ3: Population Risk Estimate Helpfulness and Re-identification Outcome Co-Occurrences}
\label{Table: MetNeedsxRe-ID Outcome}
    \centering
    \begin{tabular}{|l|l|l|l|l|l|}
    \hline
        & Not re-identified & Didn’t post & Re-identified & Outcome Unclear & Not Described \\ \hline
        No & 2 & 17 & 7 & 0 & 2 \\ \hline
        Yes & 51 & 0 & 3 & 0 & 2 \\ \hline
        Unclear & 8 & 1 & 1 & 10 & 25 \\ \hline
    \end{tabular}
\end{table}

\begin{table}[!ht]
\caption{RQ3: Ability Sub-codes and Re-identification Outcome Co-Occurrences}
\label{Table: AbilityxRe-ID Outcome}
    \centering
    \begin{tabular}{|l|l|l|l|l|l|}
    \hline
         & Not re-identified & Didn’t post & Re-identified & Outcome Unclear & Not Described \\ \hline
        Needs More Guidance & 21 & 9 & 4 & 2 & 16 \\ \hline
        Provides Balanced posting & 4 & 0 & 0 & 0 & 0 \\ \hline
    \end{tabular}
\end{table}

\begin{table}[!ht]
\caption{RQ3: Overall Impression of Population Risk Estimates and Re-identification Outcome Co-Occurrences}
\label{Table: OIxRe-ID Outcome}
    \centering
    \begin{tabular}{|l|l|l|l|l|l|}
    \hline
        & Not re-identified & Didn’t post & Re-identified & Outcome Unclear & Not Described \\ \hline
        Majority Positive & 43 & 3 & 0 & 4 & 7 \\ \hline
        Mixed & 15 & 2 & 3 & 3 & 14 \\ \hline
        Majority Negative & 3 & 9 & 7 & 1 & 6 \\ \hline
        Unclear & 0 & 4 & 1 & 2 & 2 \\ \hline
    \end{tabular}
\end{table}
    \begin{table}[!ht]
    \label{DesignCodeCoOccurrence}
    \centering
    \begin{tabular}
    {l|lllll|l}
    
    \hline
        Code 
        & Design \#1 
        & Design \#2 
        & Design \#3 
        & Design \#4 
        & Design \#5 
        & Total \\ \hline
        Explainability: Not specific enough & 5 & 4 & 8 & 9 & 5 & 31 \\ 
        Explainability: Misinterpretation   & 5 & 6 & 0 & 0 & 2 & 13 \\ 
        Explainability: Unclear Meaning     & 6 & 16 & 8 & 4 & 5 & 39 \\ \hline
        Interpretability (Con) & 3 & 3 & 3 & 3 & 5 & 17 \\ 
        Interpretability (Pro) & 2 & 1 & 0 & 0 & 0 & 3 \\ \hline
        Transparency & 5 & 3 & 3 & 3 & 3 & 17 \\ \hline
    \end{tabular}
    \caption{Summary of frequencies of iteratively developed codes for analyzing differences across PRE designs in usability. The categories are grouped around three concepts from the literature on explainable AI (XAI): explainability, interpretability, and transparency. The rightmost column tallied up the total number of occurrences of each code. \newline \newline *** Design \#1: (Raw Anonymity Score),  Design \#2 (Re-identifiability Meter), Design \#3 (Simplified Risk Level), Design \#4 (Threat-Specific Risk), Design \#5 (Risk by Disclosure)}
\end{table}

    \clearpage

\section{All Comic-board Variations}
\onecolumn

\begin{figure}
  \includegraphics[width=0.88\paperwidth]{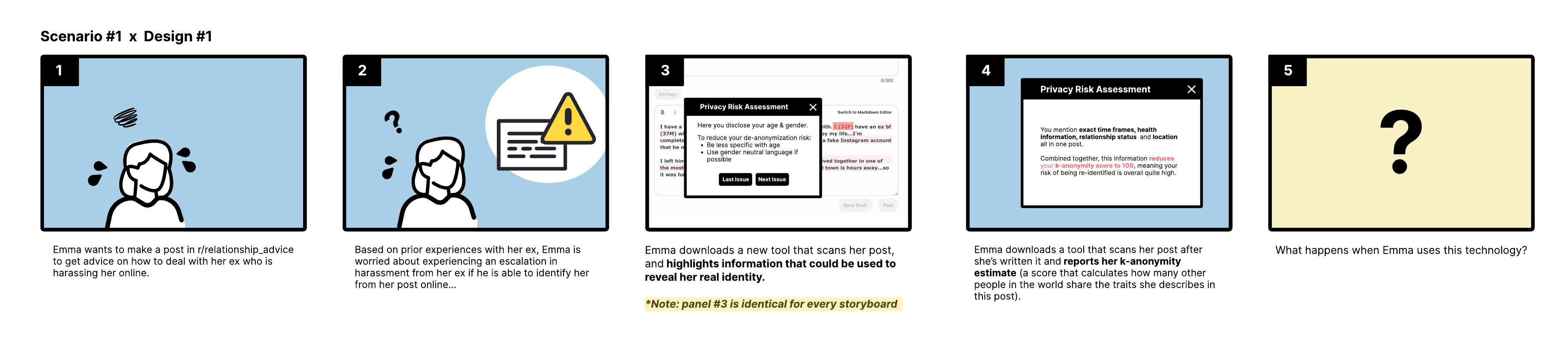}
  \caption{Scenario 1 x Design 1}
  \Description{This image depicts the comic-board combining narrative vignette scenario 1 with design 1}
  \label{}
\end{figure}

\begin{figure}
  \includegraphics[width=0.88\paperwidth]{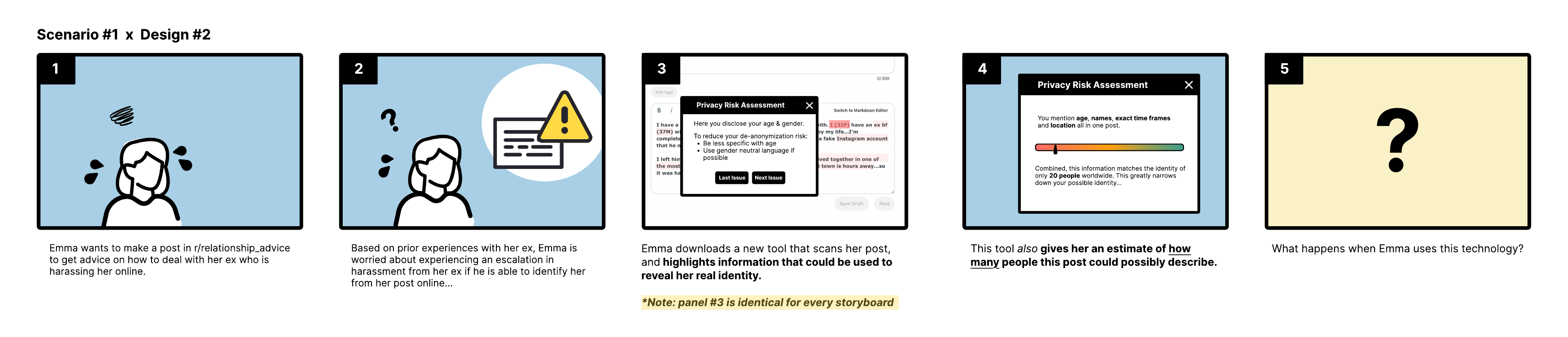}
  \caption{Scenario 1 x Design 2}
  \Description{This image depicts the comic-board combining narrative vignette scenario 1 with design 2}
\end{figure}

\begin{figure}
  \includegraphics[width=0.88\paperwidth]{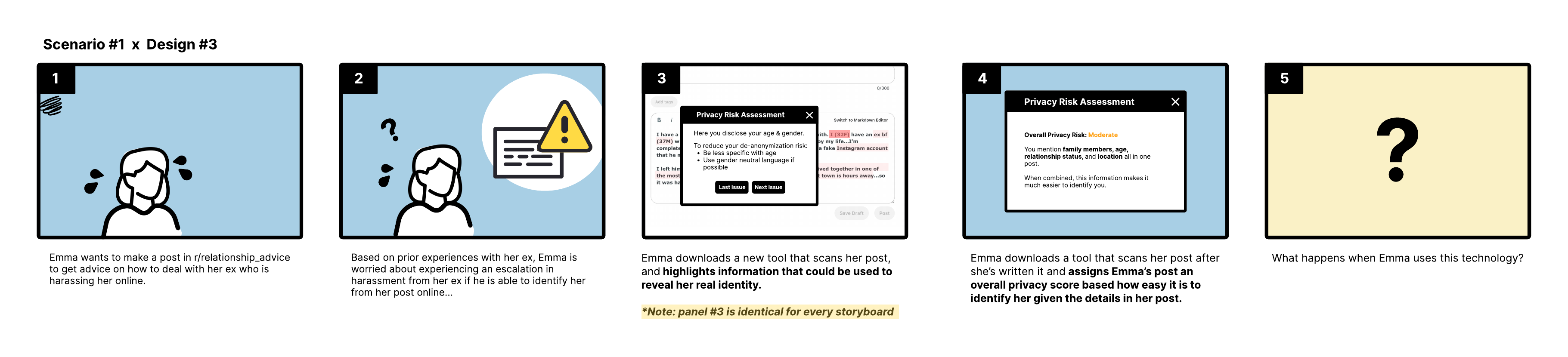}
  \caption{Scenario 1 x Design 3}
  \Description{This image depicts the comic-board combining narrative vignette scenario 1 with design 3}
\end{figure}

\begin{figure}
  \includegraphics[width=0.88\paperwidth]{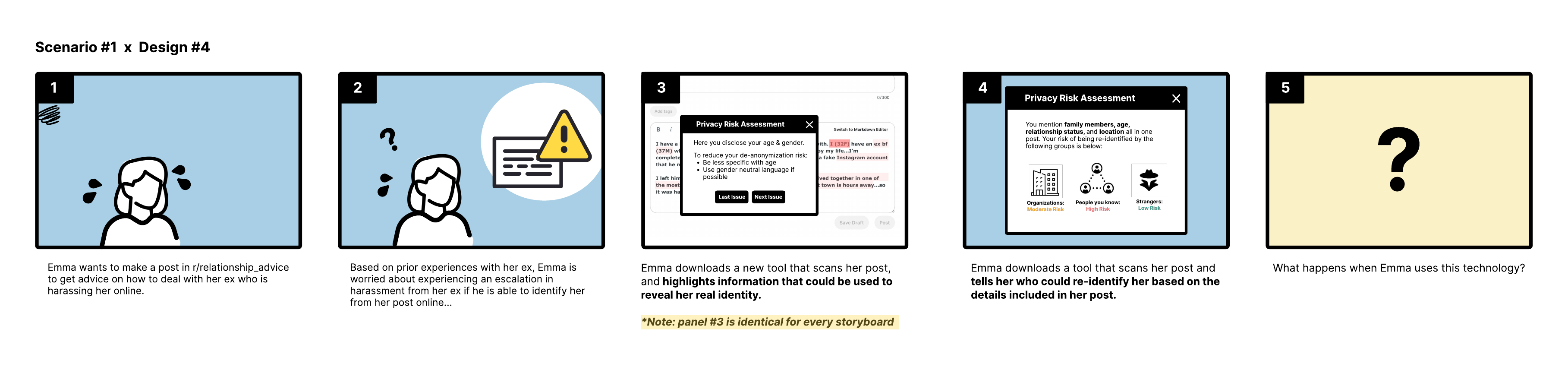}
  \caption{Scenario 1 x Design 4}
  \Description{This image depicts the comic-board combining narrative vignette scenario 1 with design 4}
\end{figure}

\begin{figure}
  \includegraphics[width=0.88\paperwidth]{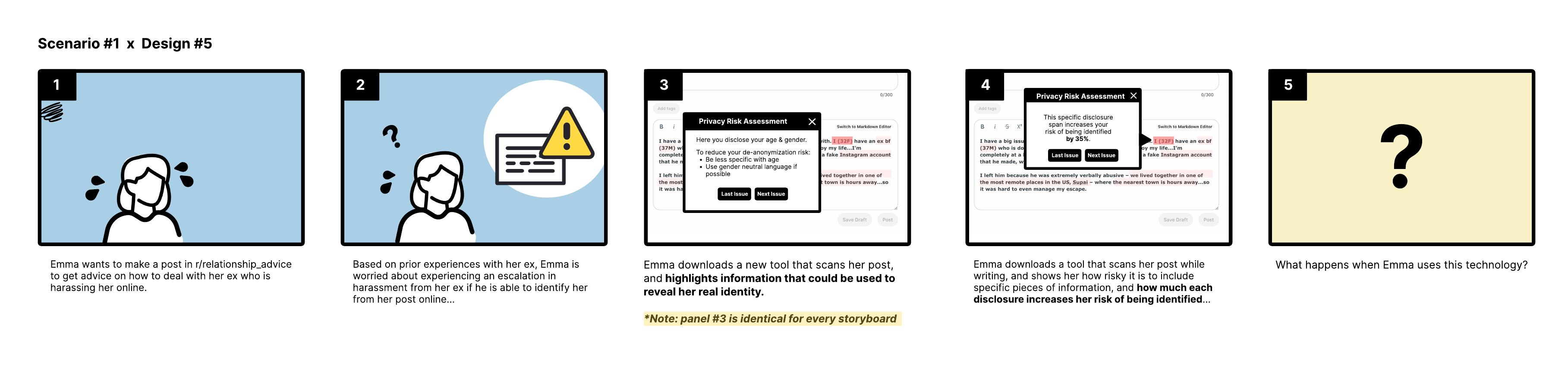}
  \caption{Scenario 1 x Design 5}
  \Description{This image depicts the comic-board combining narrative vignette scenario 1 with design 5}
\end{figure}

\begin{figure}
  \includegraphics[width=0.88\paperwidth]{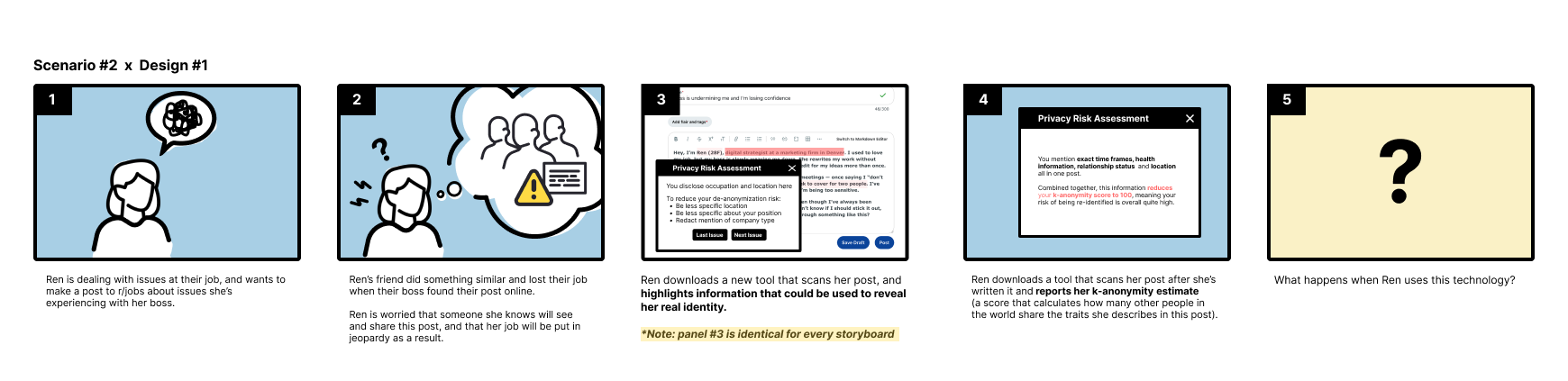}
  \caption{Scenario 2 x Design 1}
  \Description{This image depicts the comic-board combining narrative vignette scenario 2 with design 1}
\end{figure}

\begin{figure}
  \includegraphics[width=0.88\paperwidth]{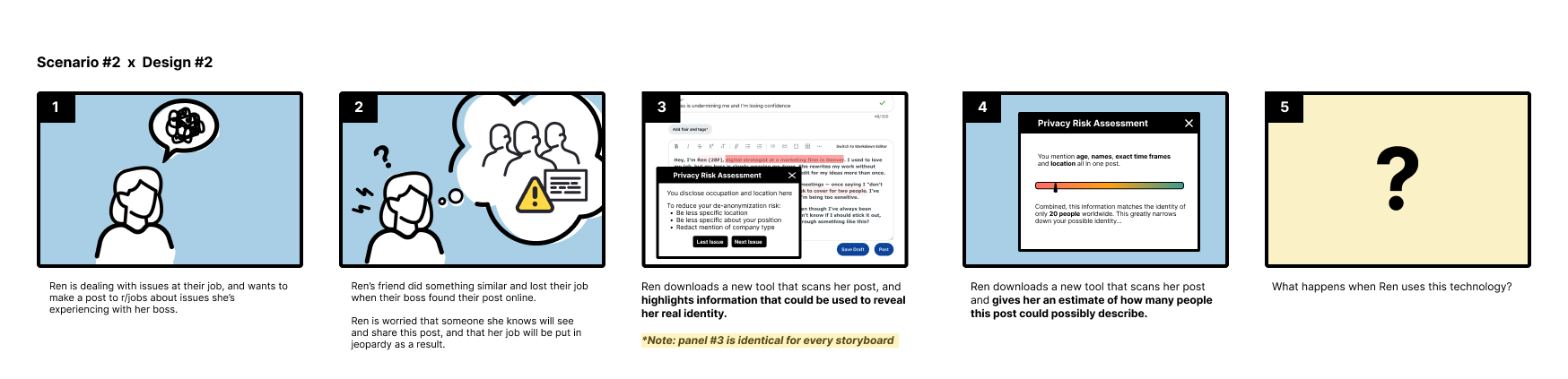}
  \caption{Scenario 2 x Design 2}
  \Description{This image depicts the comic-board combining narrative vignette scenario 2 with design 2}
\end{figure}

\begin{figure}
  \includegraphics[width=0.88\paperwidth]{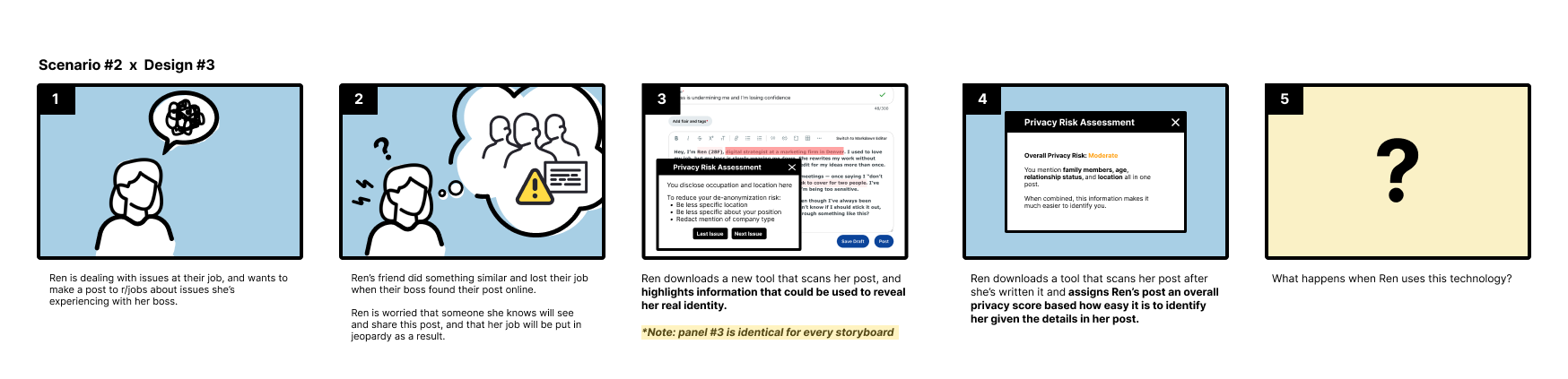}
  \caption{Scenario 2 x Design 3}
  \Description{This image depicts the comic-board combining narrative vignette scenario 2 with design 3}
\end{figure}

\begin{figure}
  \includegraphics[width=0.88\paperwidth]{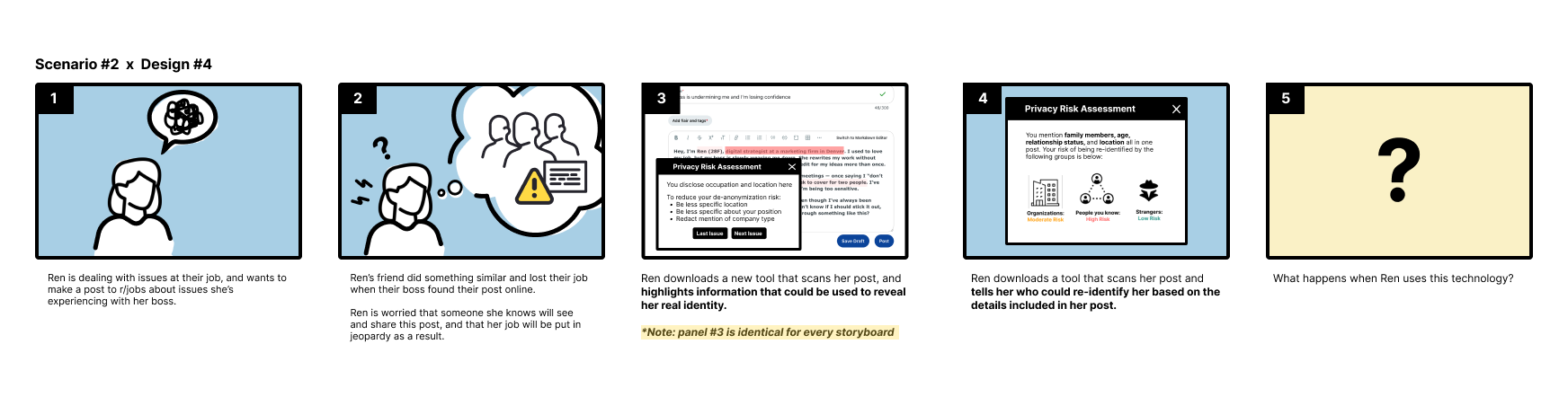}
  \caption{Scenario 2 x Design 4}
  \Description{This image depicts the comic-board combining narrative vignette scenario 2 with design 4}
\end{figure}

\begin{figure}
  \includegraphics[width=0.88\paperwidth]{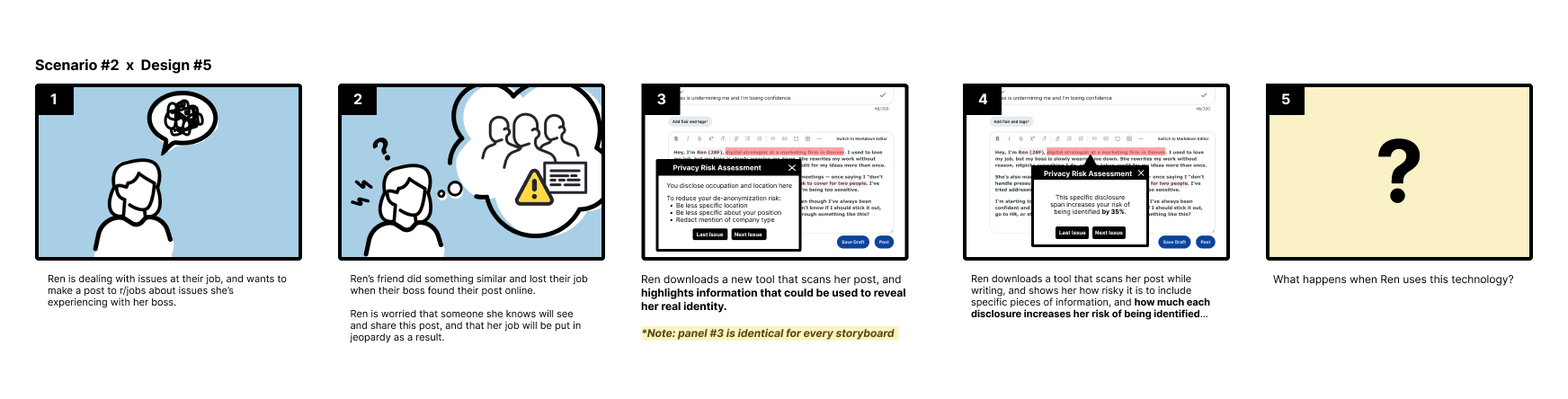}
  \caption{Scenario 2 x Design 5}
  \Description{This image depicts the comic-board combining narrative vignette scenario 2 with design 5}
\end{figure}

\begin{figure}
  \includegraphics[width=0.88\paperwidth]{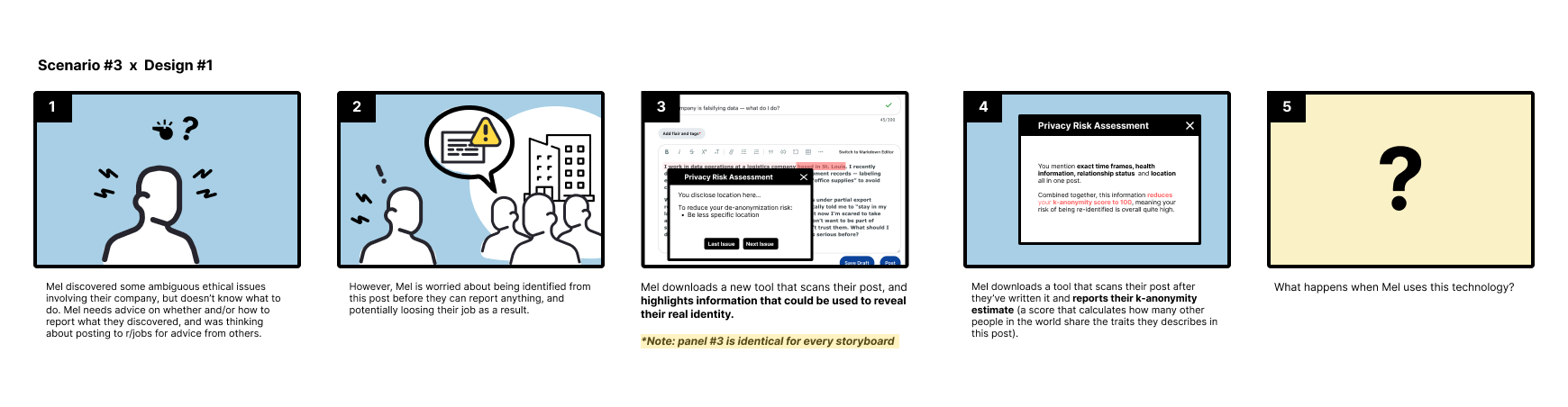}
  \caption{Scenario 3 x Design 1}
  \Description{This image depicts the comic-board combining narrative vignette scenario 3 with design 1}
\end{figure}

\begin{figure}
  \includegraphics[width=0.88\paperwidth]{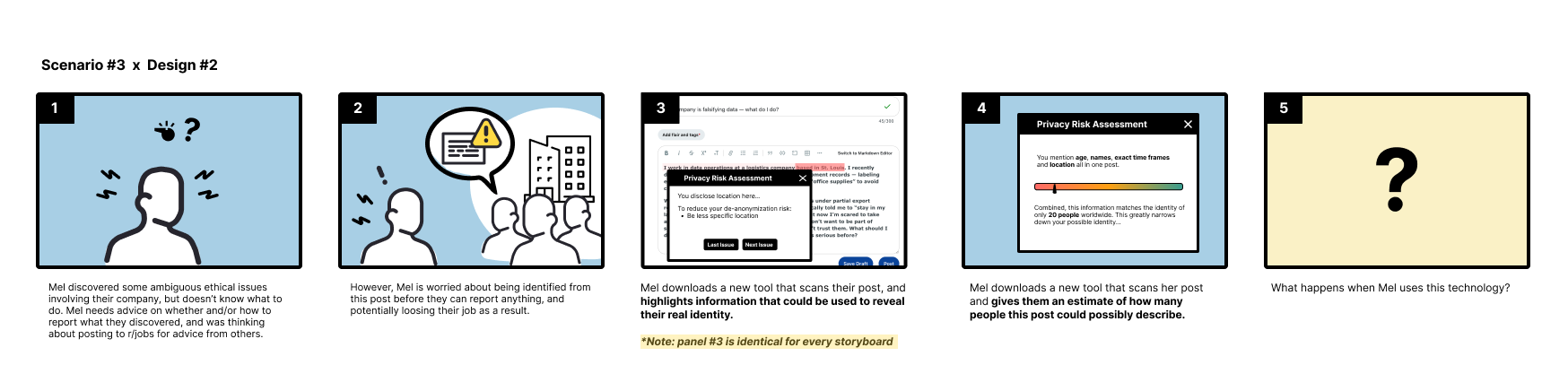}
  \caption{Scenario 3 x Design 2}
  \Description{This image depicts the comic-board combining narrative vignette scenario 3 with design 2}
\end{figure}

\begin{figure}
  \includegraphics[width=0.88\paperwidth]{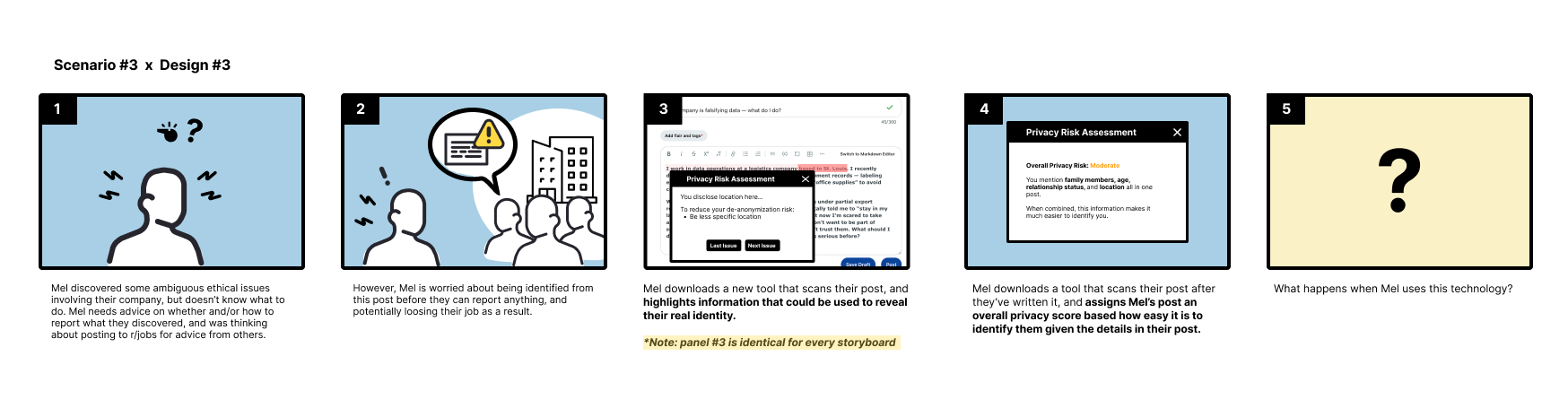}
  \caption{Scenario 3 x Design 3}
  \Description{This image depicts the comic-board combining narrative vignette scenario 3 with design 3}
\end{figure}

\begin{figure}
  \includegraphics[width=0.88\paperwidth]{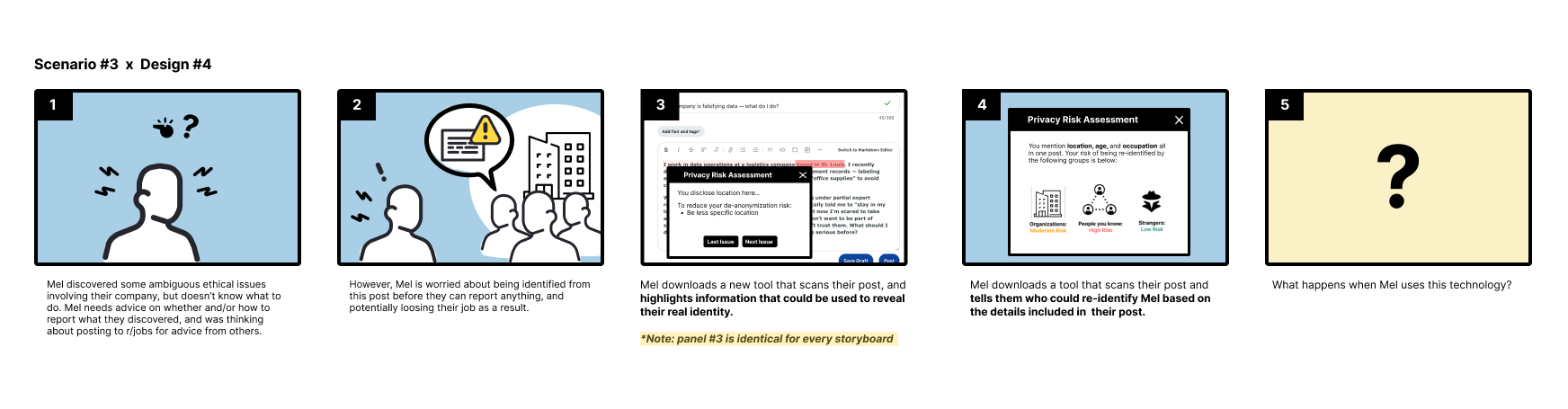}
  \caption{Scenario 3 x Design 4}
  \Description{This image depicts the comic-board combining narrative vignette scenario 3 with design 4}
\end{figure}

\begin{figure}
  \includegraphics[width=0.88\paperwidth]{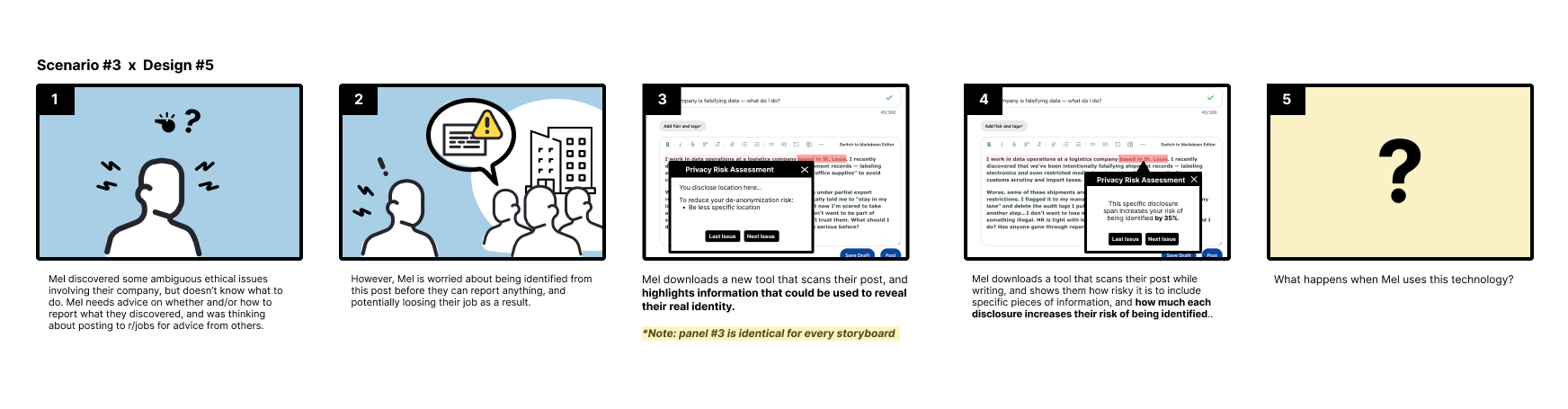}
  \caption{Scenario 3 x Design 5}
  \Description{This image depicts the comic-board combining narrative vignette scenario 3 with design 5}
\end{figure}

\begin{figure}
  \includegraphics[width=0.88\paperwidth]{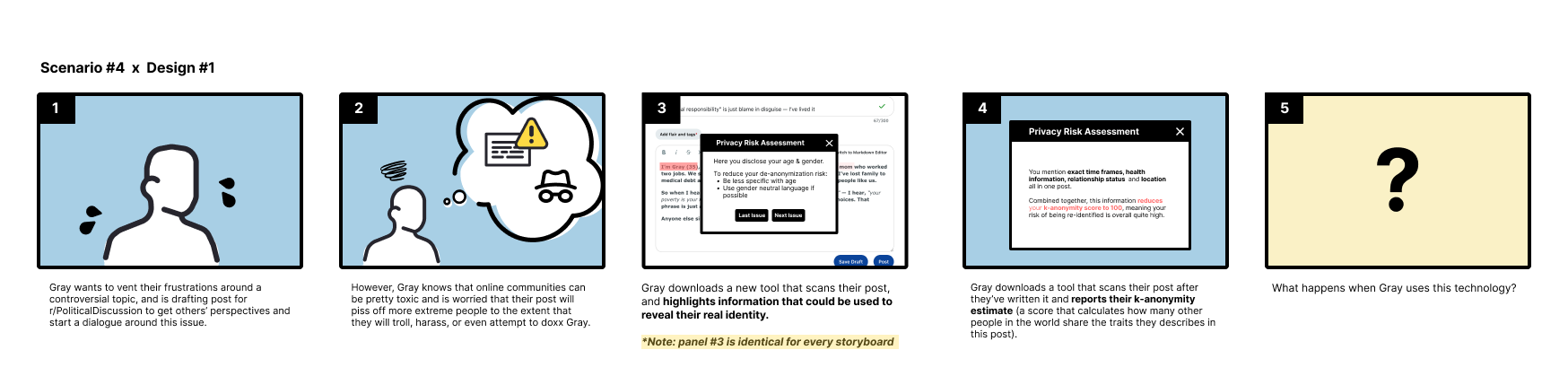}
  \caption{Scenario 4 x Design 1}
  \Description{This image depicts the comic-board combining narrative vignette scenario 4 with design 1}
\end{figure}

\begin{figure}
  \includegraphics[width=0.88\paperwidth]{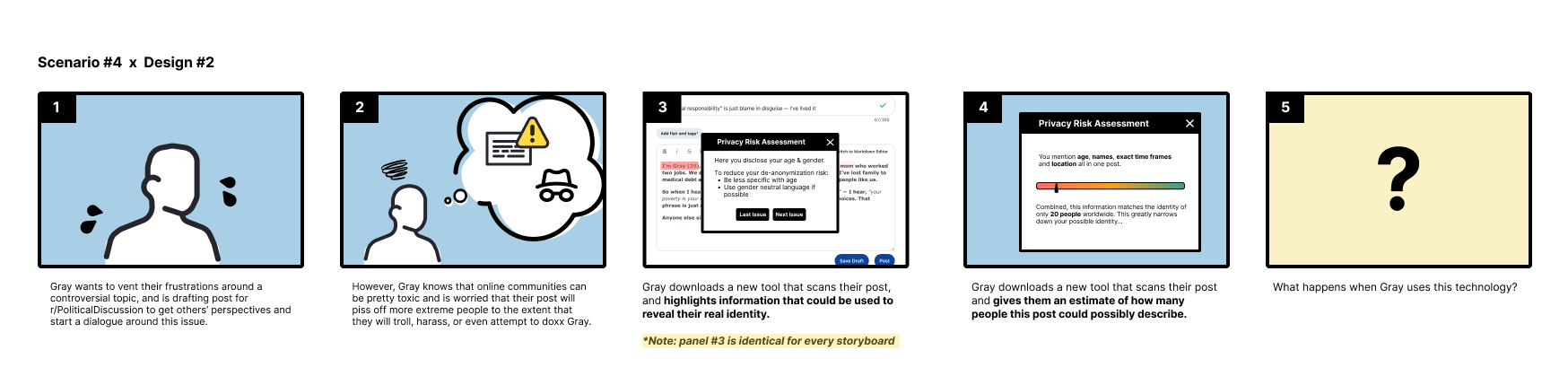}
  \caption{Scenario 4 x Design 2}
  \Description{This image depicts the comic-board combining narrative vignette scenario 4 with design 2}
\end{figure}

\begin{figure}
  \includegraphics[width=0.88\paperwidth]{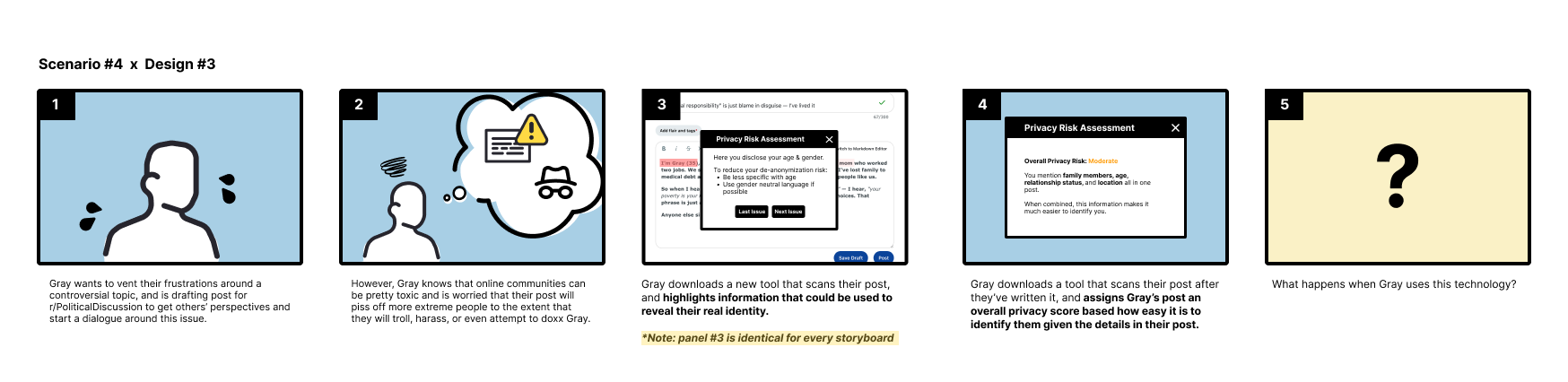}
  \caption{Scenario 4 x Design 3}
  \Description{This image depicts the comic-board combining narrative vignette scenario 4 with design 3}
\end{figure}

\begin{figure}
  \includegraphics[width=0.88\paperwidth]{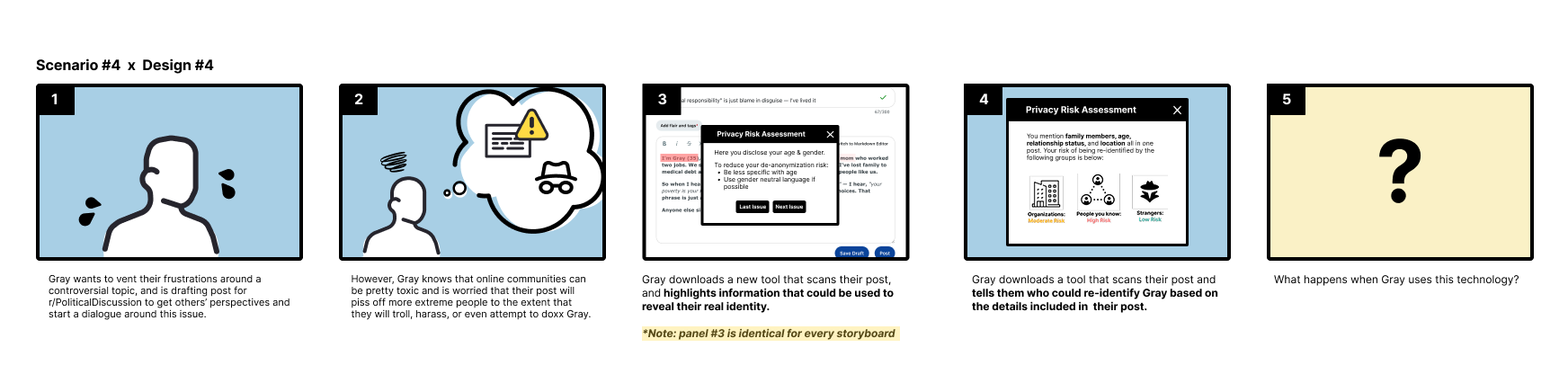}
  \caption{Scenario 4 x Design 4}
  \Description{This image depicts the comic-board combining narrative vignette scenario 4 with design 4}
\end{figure}

\begin{figure}
  \includegraphics[width=0.88\paperwidth]{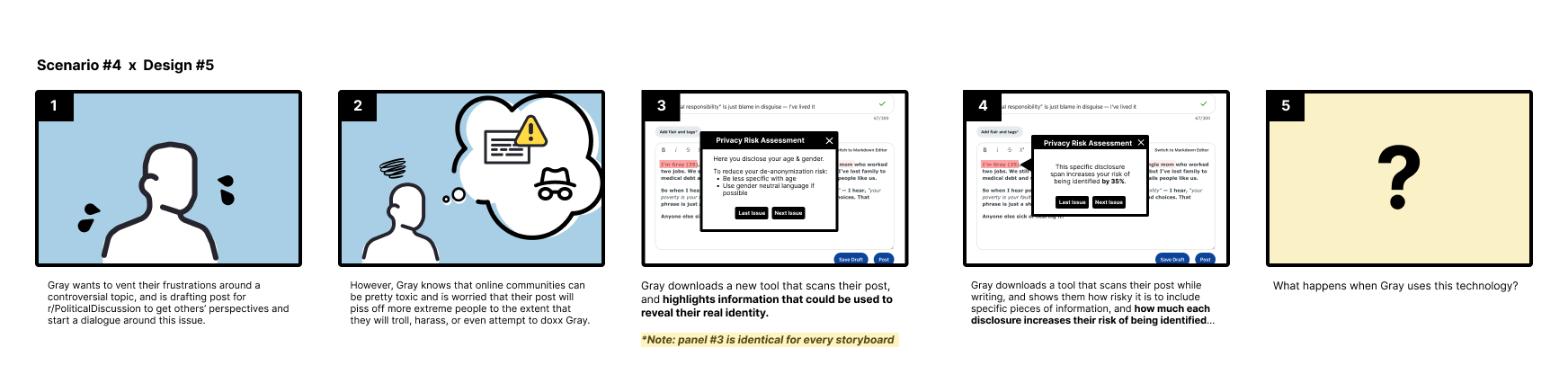}
  \caption{Scenario 4 x Design 5}
  \Description{This image depicts the comic-board combining narrative vignette scenario 4 with design 5}
\end{figure}


\end{document}